\documentclass[12pt,preprint]{aastex}

\usepackage{color}

\newcommand{\bi}{\bfseries\itshape}

\newcommand{\HI}{\mbox{H\footnotesize\,I}}
\newcommand{\HIs}{\mbox{{\scriptsize H}{\tiny\,I}}}
\newcommand{\Htwo}{\mbox{H}_{2}}
\newcommand{\Htwos}{\mbox{{\scriptsize H}}_{2}}
\newcommand{\CO}{\mbox{CO}}
\newcommand{\COs}{\mbox{\scriptsize CO}}
\newcommand{\taud}{\tau_{353}}
\newcommand{\taudmain}{\tau_{353}(\mathrm{main})}
\newcommand{\Td}{T_{\mathrm{d}}}
\newcommand{\WHI}{W_{\HIs}}
\newcommand{\WHImain}{W_{\HIs}(\mathrm{main})}
\newcommand{\NHI}{N_{\HIs}}
\newcommand{\NHImain}{N_{\HIs}(\mathrm{main})}
\newcommand{\nHI}{n_{\HIs}}
\newcommand{\NHtwo}{N_{\Htwos}}
\newcommand{\dVHI}{\mathit{\Delta}V_{\HIs}}
\newcommand{\tauHI}{\tau_{\HIs}}
\newcommand{\tauHImain}{\tau_{\HIs}(\mathrm{main})}
\newcommand{\Ts}{T_{\mathrm{s}}}
\newcommand{\WCO}{W_{\COs}}
\newcommand{\XCO}{X_{\COs}}
\newcommand{\Tbg}{T_{\mathrm{bg}}}
\newcommand{\AV}{A_{V}}

\newcommand{\valXHI}{1.823\times10^{18}}

\newcommand{\MinHIVR}{{-}49}
\newcommand{\MaxHIVR}{{+}17}

\newcommand{\HIVoneround}{{-}25}
\newcommand{\HIVtworound}{{+}12}
\newcommand{\TypicalTs}{20\,\mathrm{K}\mbox{--}40\,\mathrm{K}}
\newcommand{\TypicalDensity}{40\,\mathrm{cm^{-3}}\mbox{--}160\,\mathrm{cm^{-3}}}

\newcommand{\FittaudWCO}{\taudmain=[(1.8\pm 0.8)\times 10^{-6}]\cdot\WCO\,(\mathrm{K\,km\,s^{-1}})+[(8.4\pm 5.0)\times 10^{-6}]}
\newcommand{\DerivedXCO}{1.3\times10^{20}}
\newcommand{\DerivedNHIcoef}{1.5\times10^{26}}
\newcommand{\Mtotal}{\sim1.3\times10^{4}\,M_{\odot}}
\newcommand{\MHtwo}{\sim1.2\times10^{3}\,M_{\odot}}
\newcommand{\MHIenv}{\sim1.2\times10^{4}\,M_{\odot}}

\begin{document}


\title{$\HI$, $\CO$, and {\bi Planck}{\slash}{\bi IRAS} dust properties in the\\high-latitude-cloud complex, \object{MBM 53, 54, 55} and \object{HLCG \mbox{\boldmath $92{-}35$}}; Possible evidence for an optically thick $\HI$ envelope\\around the $\CO$ clouds}


\author{Yasuo Fukui\altaffilmark{1}, Ryuji Okamoto\altaffilmark{1}, Ryohei Kaji\altaffilmark{1}, Hiroaki Yamamoto\altaffilmark{1}, Kazufumi Torii\altaffilmark{1},\\
Takahiro Hayakawa\altaffilmark{1}, Kengo Tachihara\altaffilmark{1}, John M. Dickey\altaffilmark{2}, Takeshi Okuda\altaffilmark{1, 3},\\
Akio Ohama\altaffilmark{1}, Yutaka Kuroda\altaffilmark{1}, and Toshihisa Kuwahara\altaffilmark{1}}
%% \affil{Department of Astrophysics, Nagoya University, Chikusa-ku, Nagoya 464-8602}

\email{fukui@a.phys.nagoya-u.ac.jp}


\altaffiltext{1}{Department of Physics, Nagoya University, Chikusa-ku, Nagoya 464-8602, Japan}
\altaffiltext{2}{University of Tasmania, School of Maths and Physics, Private Bag 37, Hobart, TAS 7001, Australia}
\altaffiltext{3}{National Astronomical Observatory of Japan, 2-21-1 Osawa, Mitaka, Tokyo 181-8588, Japan}


\begin{abstract}
%We present an analysis of the $\HI$ and $\CO$ gas in conjunction with the {\it Planck}{\slash}\textit{IRAS} sub-mm{\slash}far-infrared dust properties toward the most outstanding high latitude clouds \object{MBM 53, 54, 55} and \object{HLCG $92{-}35$} at $b={-}30\,\mathrm{degree}\ \mbox{--}\ {-}45\,\mathrm{degree}$.
We present an analysis of the $\HI$ and $\CO$ gas in conjunction with the {\it Planck}{\slash}\textit{IRAS} sub-mm{\slash}far-infrared dust properties toward the most outstanding high latitude clouds \object{MBM 53, 54, 55} and \object{HLCG $92{-}35$} at $b={-}30^{\circ}$ to ${-}45^{\circ}$.
The $\CO$ emission, dust opacity at $353\,\mathrm{GHz}$ ($\taud$), and dust temperature ($\Td$) show generally good spatial correspondence.
On the other hand, the correspondence between the $\HI$ emission and the dust properties is less clear than in $\CO$.
The integrated $\HI$ intensity $\WHI$ and $\taud$ show a large scatter with a correlation coefficient of $\sim 0.6$ for a $\Td$ range from $16\,\mathrm{K}$ to $22\,\mathrm{K}$.
We find however that $\WHI$ and $\taud$ show better correlation for smaller ranges of $\Td$ every $0.5\,\mathrm{K}$, generally with a correlation coefficient of $0.7\mbox{--}0.9$.
We set up a hypothesis that the $\HI$ gas associated with the highest $\Td\geq21.5\,\mathrm{K}$ is optically thin, whereas the $\HI$ emission is generally optically thick for $\Td$ lower than $21.5\,\mathrm{K}$.
We have determined a relationship for the optically thin $\HI$ gas between atomic hydrogen column density and $\taud$, $\NHI\,(\mathrm{cm^{-2}})=(\DerivedNHIcoef)\cdot\taud$, under the assumption that the dust properties are uniform, and applied it to estimate $\NHI$ from $\taud$ for the whole cloud.
$\NHI$ was then used to solve for $\Ts$ and $\tauHI$ over the region. The result shows that the $\HI$ is dominated by optically thick gas having low spin temperature of $\TypicalTs$ and density of $\TypicalDensity$.
The $\HI$ envelope has a total mass of $\MHIenv$, an order of magnitude larger than that of the $\CO$ clouds.
\textcolor{black}{The $\HI$ envelope properties derived by this method do not rule out a mixture of $\HI$ and $\Htwo$ in the dark gas, but we present indirect evidence that most of the gas mass is in the atomic state.
}
\end{abstract}





\section{Introduction}

The neutral interstellar medium (ISM) consists of $\HI$, $\Htwo$, and possibly ``dark gas'', which is believed to be undetectable in line emission of $\HI$ or $\CO$ \citep{2005Sci...307.1292G,2011A&A...536A..19P}.
The $\HI$ emission at $21\,\mathrm{cm}$ wavelength and the $\CO$ emission at $2.6\,\mathrm{mm}$ offer tools to probe atomic and molecular hydrogen.
The $\CO$ clouds have low kinetic temperature $10\,\mathrm{K}\mbox{--}100\,\mathrm{K}$ with density above $1000\,\mathrm{cm^{-3}}$ as derived by analyses of multi-$J$ $\CO$ transitions \citep[e.g.][]{1990A&A...234..469C}.
The $\CO$ intensity is converted into molecular column density by an $\XCO$ factor, which is empirically determined by assuming dynamical equilibrium of the $\CO$ clouds, dust properties measured in extinction \textcolor{black}{and{\slash}or far-infrared} emission, and comparison with gamma rays created by proton-proton collisions via neutral pion decay \citep[e.g.,][]{2007prpl.conf...81B,2010ARA&A..48..547F,2013ARA&A..51..207B}.
On the other hand, the physical parameters of the $\HI$ gas have been rather ambiguous, mainly because the $\HI$ emission has a single observed quantity, intensity as a function of velocity, for two independent variables spin temperature $\Ts$ and $\HI$ optical depth $\tauHI$, not allowing us to determine each of these observationally.
The $\HI$ consists of warm and cold components \citep[for a review see][]{2009ARA&A..47...27K,1990ARA&A..28..215D}.
The mass of the $\HI$ gas having spin temperature higher than $\sim 100\,\mathrm{K}$ is measurable at reasonably high accuracy under the optically thin approximation, while the cold component having spin temperature of $20\,\mathrm{K}\mbox{--}80\,\mathrm{K}$ is measured by comparing the absorption and emission $\HI$ profiles only where background continuum sources are available \citep{2003ApJ...585..801D,2003ApJ...586.1067H}.
It is notable that some recent works on TeV gamma-ray supernova remnants show that the cold $\HI$ gas which is not detectable in $\CO$ at $1\,\sigma$ integrated intensity level of $\sim1\,\mathrm{K\,km\,s^{-1}}$ \citep{2012ApJ...746...82F} is responsible for the gamma-rays via collision between cosmic-ray protons and interstellar protons \citep{2012ApJ...746...82F,2013ASSP...34..249F,2014ApJ...788...94F}.
However, the observed sample of cold $\HI$ is limited and the cold $\HI$ still needs further effort to better constrain its physical properties.

Recent progress in measuring the infrared and sub-mm emission over the whole sky is remarkable.
These observations are strongly motivated by the aim to precisely measure the cosmic microwave background (CMB) radiation and its polarization.
In particular, the \textit{Planck} collaboration released the whole sky distribution of dust temperature $\Td$ and dust opacity $\taud$ at $5\,\mathrm{arcmin}$ resolution which have been achieved through sensitive measurements of the dust emission above $100\,\mathrm{GHz}$ in the Galactic foreground.
Five papers of the \textit{Planck} collaboration presented comparisons between $\CO$ and dust emission \citep{2011A&A...536A..19P,2011A&A...536A..21P,2011A&A...536A..22P,2011A&A...536A..23P,2011A&A...536A..25P}.
These studies derived basic physical relationships between the neutral gas and the dust emission and concluded that the dark gas \citep{2005Sci...307.1292G} occupies $15\,\%$ of the interstellar medium (ISM), where the $\HI$ emission was assumed to be optically thin.
Similar studies on $21\,\mathrm{cm}$ $\HI$ emission include \citet{2012ApJ...748...75L} and \citet{2011A&A...536A..24P}.

The high latitude molecular clouds \object{MBM 53, 54, 55} and \object{HLCG $92{-}35$} are one of the largest cloud systems in the sky at Galactic latitude of ${-}30\,\mathrm{degree}\ \mbox{--}\ {-}50\,\mathrm{degree}$ with only minor background components \citep{1985ApJ...295..402M}.
The distance of the cloud system is determined to be $150\,\mathrm{pc}$ \citep{1989ApJ...346..232W}.
\citet{2003ApJ...592..217Y} used the NANTEN $4\,\mathrm{m}$ telescope to observe the $J{=}1\mbox{--}0$  transitions of $^{12}\CO$, $^{13}\CO$, and $\mathrm{C}^{18}\mathrm{O}$ and mapped the molecular distribution.
The velocity of the \object{MBM 53, 54, 55} and \object{HLCG $92{-}35$} clouds is mainly in range from ${-}11\,\mathrm{km\,s^{-1}}$ to $0\,\mathrm{km\,s^{-1}}$ for $\CO$.
The half-power beam widths were $2.6\,\mathrm{arcmin}\mbox{--}2.7\,\mathrm{arcmin}$ and their grid spacings are $4$ or $8\,\mathrm{arcmin}$ for the $^{12}\CO$, $2\,\mathrm{arcmin}$ for the $^{13}\CO$ and the $\mathrm{C}^{18}\mathrm{O}$.
They also identified associated $\HI$ gas at $36\,\mathrm{arcmin}$ resolution and discussed that the $\HI$ gas is being converted into $\Htwo$ under the dynamical effect of a nearby shell driven by stellar winds.
\textcolor{black}{
\citet{1994ApJ...434..162G} studied and identified the associated $\HI$, while their coverage was limited to ${\sim}1/2$ of the area studied by \citet{2003ApJ...592..217Y}.
}
For reference, this region is rich in high latitude clouds including a far-infrared looplike structure in Pegasus, and \citet{2006ApJ...642..307Y} observed this loop in $\CO(J{=}1\mbox{--}0)$ lines.
\object{MBM 53, 54, 55} and \object{HLCG $92{-}35$} clouds show no sign of active star formation and are an ideal target to study detailed ISM properties under the general interstellar radiation field.
The regions of the \object{MBM 53, 54, 55} and \object{HLCG $92{-}35$} clouds and the Pegasus loop have also been used to test $\CO$ component separation in the analysis of the cosmic microwave background emission \citep{2014ApJ...780...13I}.

In order to better understand the detailed physical properties of the local ISM, we have undertaken new $\CO(J{=}1\mbox{--}0)$ observations of \object{MBM 53, 54, 55} and \object{HLCG $92{-}35$} in the on-the-fly mode with the $2.6\,\mathrm{arcmin}$ beam of the NANTEN2 $4\,\mathrm{m}$ telescope as part of the \underline{NA}NTEN \underline{S}uper \underline{$\CO$} survey (=NASCO), which is aiming to cover $70\,\%$ of the sky.
By using the $\HI$ dataset at $4\,\mathrm{arcmin}$ resolution from the GALFA archive data, we have compared the $\CO$ and $\HI$ with the \textit{Planck}{\slash}\textit{IRAS} dust properties.
In this paper we present the first results from the detailed comparison. Section 2 describes the observations, Section 3 gives the results and analysis, and Section 4 presents conclusions.
Another paper which deals with the dust emission covering the whole sky will be published separately \citep[][]{2014arXiv1403.0999F} and is complementary to the present paper.

\section{Observational datasets}

In this study we used NANTEN2 $\CO$ data, GALFA $\HI$ data, the dust emission data measured by \textit{Planck}{\slash}\textit{IRAS}, and the $1.4\,\mathrm{GHz}$ radio continuum data.
All the data are gridded to $1\,\mathrm{arcmin}$, smaller than the angular resolutions of these datasets.

\subsection{$\CO$ data}

$^{12}\CO(J{=}1\mbox{--}0)$ and $^{13}\CO(J{=}1\mbox{--}0)$ emission were obtained by the NANTEN2 $4\,\mathrm{m}$ millimeter{\slash}submillimeter telescope of Nagoya University, which is installed at an altitude of $4865\,\mathrm{m}$ in northern Chile.
Observations of the $^{12}\CO(J{=}1\mbox{--}0)$ and $^{13}\CO(J{=}1\mbox{--}0)$ lines were simultaneously conducted in several sessions from 2011 November to 2013 January, and cover about $75\,\mathrm{deg}^{2}$.
Only observational parameters of $^{12}\CO(J{=}1\mbox{--}0)$ are described below.
The front end was a $4\,\mathrm{K}$ cooled Nb Superconductor-Insulator-Superconductor (SIS) mixer receiver, and the double-sideband (DSB) system temperature was $\sim 200\,\mathrm{K}$ including the atmosphere toward the zenith.
The spectrometer was a digital Fourier transform spectrometer (DFS) with 16,384 channels, providing velocity resolution of $0.2\,\mathrm{km\,s^{-1}}$.
All observations were conducted by using an on-the-fly (OTF) mapping technique with a $2.6\,\mathrm{arcmin}$ main beam at a $1\,\mathrm{arcmin}$ data sampling, and each spatial area of $1\,\mathrm{degree}$ by $1\,\mathrm{degree}$ was mapped several times in different scanning directions to reduce scanning effects by basket-weave technique \citep{1988A&A...190..353E}.
These OTF data were reduced into {\sc fits} data cubes with the original {\sc idl} software.
The rms noise fluctuations in $^{12}\CO(J{=}1\mbox{--}0)$ spectra after basket-weave technique were $\sim 1.2\,\mathrm{K}$ per channel in the main beam temperature, $T_{\mathrm{mb}}$.
The $\CO$ data is smoothed to $5\,\mathrm{arcmin}$ spatial resolution to be compared with $\HI$ and \textit{Planck}{\slash}\textit{IRAS} datasets.
The final rms noise fluctuations are $\sim 0.6\,\mathrm{K}$ per channel in a $5\,\mathrm{arcmin}$ smoothed beam.
The pointing was checked regularly using planets, and the applied corrections were always smaller than $20\,\mathrm{arcsec}$.
The standard sources M17 SW and Ori KL were observed for intensity calibration.
The $\CO$ intensity is integrated over a velocity range from $-12\,\mathrm{km\,s^{-1}}$ to ${+}2\,\mathrm{km\,s^{-1}}$.

\subsection{$\HI$ data}

Archival datasets of the GALFA-$\HI$ survey are used \citep{2011ApJS..194...20P}.
The survey was conducted on the Arecibo Observatory $305\,\mathrm{m}$ telescope and the angular resolution of the data is $4\,\mathrm{arcmin}$.
The rms noise fluctuations of the $\HI$ data are $\sim0.25\,\mathrm{K}$ per channel in $T_{\mathrm{mb}}$.
The $\HI$ data are also smoothed to $5\,\mathrm{arcmin}$ resolution, and the rms noise fluctuations of the final $\HI$ spectra are $0.2\,\mathrm{K}$ per channel for the $5\,\mathrm{arcmin}$ beam.
The $\HI$ intensity \textcolor{black}{is spread over} a velocity range from $\MinHIVR\,\mathrm{km\,s^{-1}}$ to $\MaxHIVR\,\mathrm{km\,s^{-1}}$, while most of the $\HI$ emission is \textcolor{black}{between} $\HIVoneround\,\mathrm{km\,s^{-1}}$ \textcolor{black}{and} $\HIVtworound\,\mathrm{km\,s^{-1}}$.

\subsection{\textit{Planck}{\slash}\textit{IRAS} data}

Archival datasets of the optical depth at $353\,\mathrm{GHz}$ ($\taud$) and dust temperature ($\Td$) were obtained by fitting $353$, $545$, and $857\,\mathrm{GHz}$ data of the first 15 months observed by the \textit{Planck} satellite and $100\,\mathrm{micron}$ data measured by the \textit{IRAS} satellite.
Both of these data have angular resolutions of $5\,\mathrm{arcmin}$.
For details, see the \textit{Planck} Legacy Archive (PLA) explanatory supplement \citep{PlanckExplanatorySupplement}.

\subsection{Radio continuum data}
As the $1.4\,\mathrm{GHz}$ radio continuum background we use the CHIPASS data \citep{2014PASA...31....7C} for the region at $\delta(\mathrm{J2000})<25^{\circ}$ and the data described in \citet{1982A&AS...48..219R} and \citet{1986A&AS...63..205R} for the region at $\delta(\mathrm{J2000})\geq25^{\circ}$.
Their spatial resolutions are $14.4\,\mathrm{arcmin}$ and $35\,\mathrm{arcmin}$, respectively.
Although these data have lower angular resolutions than the rest analyzed in the study, the peak radiation temperature of the continuum data shows small variations from $\sim3.4\,\mathrm{K}$ to $\sim4.9\,\mathrm{K}$ except for the area within $\sim30\,\mathrm{arcmin}$ of the quasar 3C\,454.3, which happens to be located near the $\CO$ clouds \citep[e.g.,][]{2011ApJ...733L..26A}.
We used these lower resolution data by excluding the area around 3C\,454.3.

\subsection{Masking}
In the present analysis, the local interstellar medium toward the cloud system is approximated by a single layer.
This is consistent with the velocity of the $\HI$ \textcolor{black}{being} peaked at around $0\,\mathrm{km\,s^{-1}}$\textcolor{black}{,} and generally \textcolor{black}{a} single \textcolor{black}{velocity} component is dominant.
In order to eliminate possible sources of contamination, we further applied the following masking.
These masked areas are listed in Table 1 (see also Figure~\ref{tstauhimap}).
\begin{itemize}
\item[(1)] \textcolor{black}{The areas showing $\CO$ emission \textcolor{black}{higher} than $1.1\,\mathrm{K\,km\,s^{-1}}$ ($1\sigma$) are masked in order to exclude the contribution of the $\CO$-emitting molecular gas.}
\item[(2)] \textcolor{black}{The two regions where \textcolor{black}{there is} a secondary $\HI$ peak outside the velocity range from $-25\,\mathrm{km\,s^{-1}}$ to $+12\,\mathrm{km\,s^{-1}}$ are masked at $(\ell, b)\sim(81\fdg5, -33\fdg0)$, $(85\fdg6, -33\fdg8)$. Generally, these secondary peaks show intensity comparable to that of the main $\HI$ peak. }
\item[(3)] Regions with high $\Td$ indicating localized heating by stars etc. are masked by referring to the \textit{IRAS} point source catalog etc. The most extended region of such high $\Td$ is located toward 3C\,454.3, although the origin of the high $\Td$ is not known.
\item[(4)] The six regions where the $\HI$ data are missing are masked.
\end{itemize}

\section{Results}

\begin{figure}
\epsscale{1.0}
\plotone{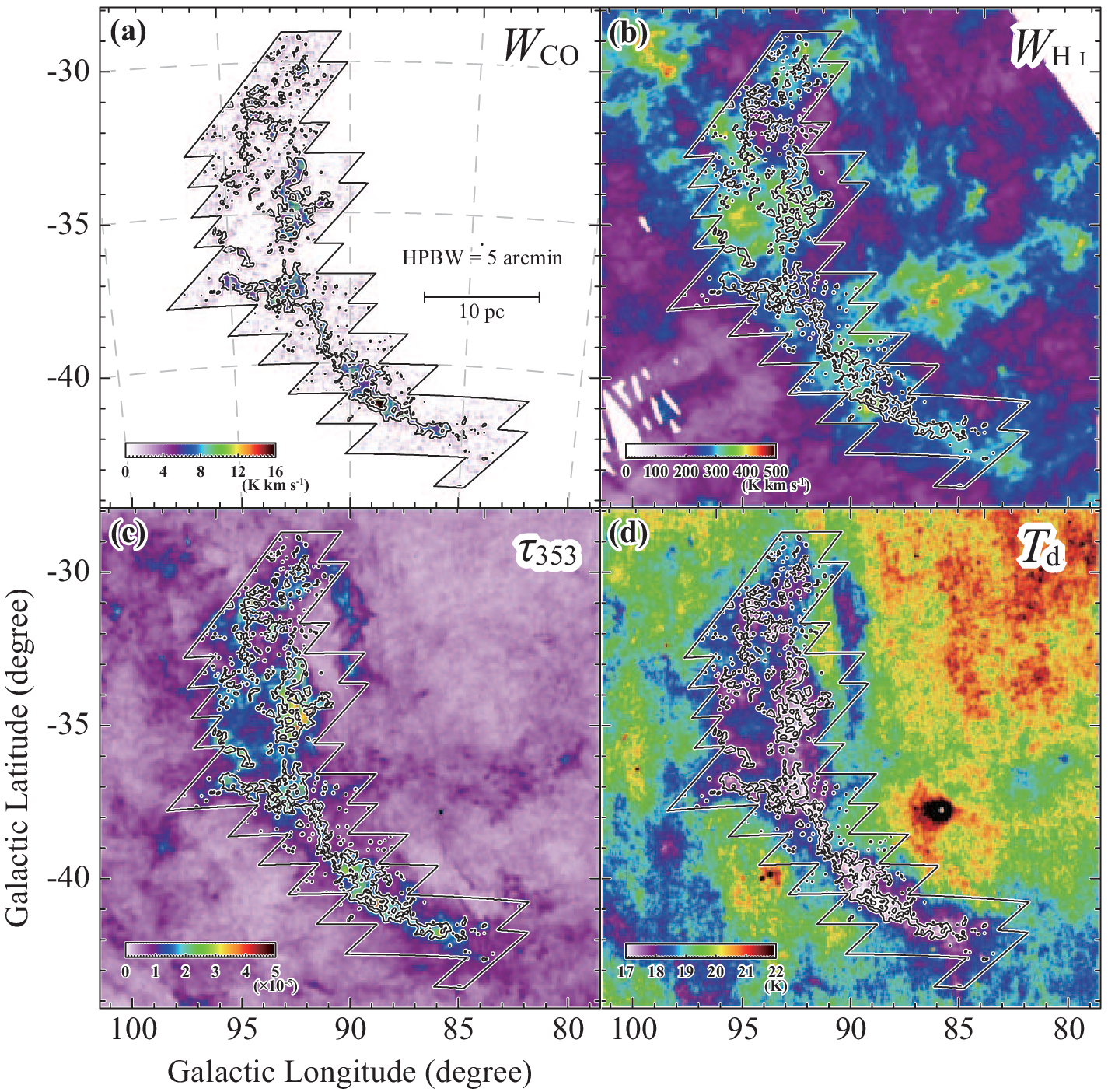}
\caption{\textcolor{black}{{\bf (a)} Velocity-integrated intensity map of $^{12}\CO(J{=}1\mbox{--}0)$ in the galactic coordinates. The integrated LSR-velocity range is from ${-}12\,\mathrm{km\,s^{-1}}$ to ${+}2\,\mathrm{km\,s^{-1}}$. The contours are drawn every $7.9\,\mathrm{K\,km\,s^{-1}}$ from $3.4\,\mathrm{K\,km\,s^{-1}}$. The bounding box shows the observed area. {\bf (b)}, {\bf (c)}, and {\bf (d)} are spatial distributions of velocity-integrated intensity of the $\HI$ ($\WHI$), the optical depth at $353\,\mathrm{GHz}$ ($\taud$), and the temperature of the cold dust ($\Td$), respectively. The contours of $\CO$ and the bounding boxes are the same as in {\bf (a)}. The integrated velocity range of the $\WHI$ in {\bf (b)} is from $\MinHIVR\,\mathrm{km\,s^{-1}}$ to $\MaxHIVR\,\mathrm{km\,s^{-1}}$. The effective beam sizes are $5\,\mathrm{arcmin}$ for the 4 panels.}}
\label{maps}
\end{figure}

Figure~\ref{maps} shows four panels which present $\CO$, $\HI$ and \textit{Planck}{\slash}\textit{IRAS} dust properties.
Figure~\ref{maps}(a) shows the $\CO(J{=}1\mbox{--}0)$ velocity integrated intensity \textcolor{black}{distribution}, $\WCO\,(\mathrm{K\,km\,s^{-1}})$.
\textcolor{black}{
Figure~\ref{maps}(b) shows an overlay of $\HI$ and $\CO$ where we confirm the associated $\HI$ in the previous works \citep{1994ApJ...434..162G,2003ApJ...592..217Y}.
$\WHI$ ($\mathrm{K\,km\,s^{-1}}$) is the intensity of $\HI$ integrated from $\MinHIVR\,\mathrm{km\,s^{-1}}$ to $\MaxHIVR\,\mathrm{km\,s^{-1}}$.
Figure~\ref{maps}(c) presents an overlay of $\taud$ and $\CO$.
The comparison shows good correspondence between them within the $\CO$ contours, whereas the correspondence of $\HI$ with $\taud$ seems less clear than in the case of $\CO$.
We note that $\taud$ is significantly extended beyond the lowest $\CO$ contour $3.4\,\mathrm{K\,km\,s^{-1}}$ ($3\,\sigma$).
}
Figure~\ref{maps}(d) shows an overlay of $\Td$ and $\CO$.
The lowest $\Td\sim16\,\mathrm{K}$ is clearly associated with $\CO$, and $\Td$ increases up to ${\sim}22\,\mathrm{K}$ outside the $\CO$ clouds.

\begin{figure}
\epsscale{0.8}
\plotone{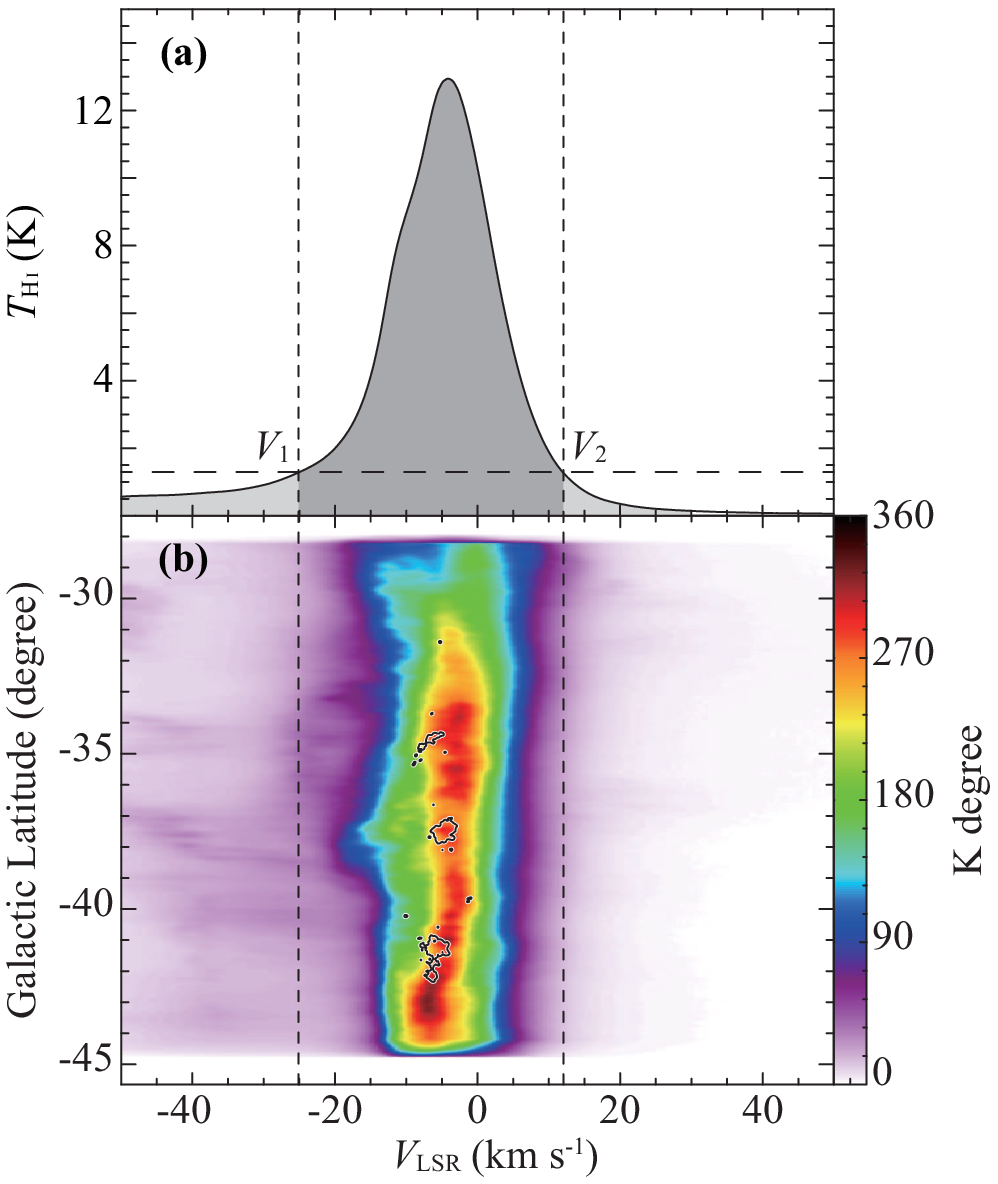}
\caption{{\bf (a)} Average spectrum of the $\HI$ in the region shown in Figure~\ref{maps}(d)\textcolor{black}{, but the masking described in Section 2.5. are applied}. $T_{\mathrm{mb}}$ at $V_{1}=\HIVoneround\,\mathrm{km\,s^{-1}}$ and at $V_{2}=\HIVtworound\,\mathrm{km\,s^{-1}}$ correspond to $10\,\%$ level of the peak. {\bf (b)} Latitude-velocity diagram of $\HI$ (color image) and $\CO$ (contours) integrated over the longitude range in Figure~\ref{maps}. The integrating range are from $\ell\sim80^{\circ}$ to $\sim100^{\circ}$. The level of the $\CO$ contours is $\sim3\,\mathrm{K\,degree}$. $89\,\%$ of the $\HI$ emission is concentrated in the velocity range from $-25\,\mathrm{km\,s^{-1}}$ to $+12\,\mathrm{km\,s^{-1}}$.}
\label{vrange}
\end{figure}

\textcolor{black}{
Figure~\ref{vrange}(a) where masking in Section 2.5. is applied shows an average $\HI$ spectrum over the region in Figure~\ref{maps}, and Figure~\ref{vrange}(b), the $\HI$ integrated intensity over the same longitude range, where $^{12}\CO(J{=}1\mbox{--}0)$ intensity is superposed by black contours.
}
Figure~\ref{vrange}(a) indicates that most of the $\HI$ emission is in \textcolor{black}{the} range from $\HIVoneround\,\mathrm{km\,s^{-1}}$ to $\HIVtworound\,\mathrm{km\,s^{-1}}$ which is determined by the $10\,\%$ level of the average $\HI$ spectrum and the $\HI$ emission from $\HIVoneround\,\mathrm{km\,s^{-1}}$ to $\HIVtworound\,\mathrm{km\,s^{-1}}$ has a single component, which includes $89\,\%$ of the $\HI$ integrated intensity.
The remaining $11\,\%$ of the $\HI$ emission is mainly \textcolor{black}{on} the negative velocity \textcolor{black}{side} at $V_{\mathrm{LSR}}<-25\,\mathrm{km\,s^{-1}}$.
In order to eliminate possible contamination of the $\HI$ at $V_{\mathrm{LSR}}<-25\,\mathrm{km\,s^{-1}}$ we shall adopt the $\HI$ \textcolor{black}{velocity range} from $-25\,\mathrm{km\,s^{-1}}$ to $+12\,\mathrm{km\,s^{-1}}$ in the following analysis and call this the main $\HI$ cloud.
This velocity range includes the $\HI$ which \textcolor{black}{has been} suggested to be associated with the $\CO$ clouds \citep{1994ApJ...434..162G,2003ApJ...592..217Y}, and more details are given in Appendix A on the $\HI$ velocity channel distributions (Figure~\ref{chmaps}).

\textcolor{black}{
In the following, $\taud$ and $\WHI$ denote those \textcolor{black}{quantities} for the whole velocity range and $\taudmain$ and $\WHImain$ for the main $\HI$ cloud only.
The excluded velocity ranges from the main $\HI$ cloud have very weak $\HI$ intensity of $\lesssim30\,\mathrm{K\,km\,s^{-1}}$ (Figure~\ref{vrange}(b)) and it is likely that this $\HI$ is optically thin.
$\WHImain$ is calculated from $\WHI$ by subtracting the $\HI$ integrated intensity at $V_{\mathrm{LSR}}<-25\,\mathrm{km\,s^{-1}}$ and $V_{\mathrm{LSR}}>+12\,\mathrm{km\,s^{-1}}$.
By adopting the relationship between $\taud$ and $\WHI$ in the optically-thin limit (see below), $\taudmain$ is calculated from $\taud$ by subtracting \textcolor{black}{a fraction of $\taud$ in proportion to $[\WHI-\WHImain]/\WHI$, which corrects for the emission due to the velocity range excluded from $\HI$ main.}
The details of the subtraction are given in Appendix B.
}

\textcolor{black}{
Figures~\ref{tau353whi1}(a) and \ref{tau353whi2}(a) show scatter plots between $\WHI$ and $\taud$ for the main $\HI$ cloud and for the whole $\HI$ velocity span, respectively, in the region with masking (Section 2.5.).
We note that Figures~\ref{tau353whi1} and \ref{tau353whi2} show the same trend\textcolor{black}{,} as described below, indicating that the subtraction has a minor effect \textcolor{black}{on} the results as expected from the weakness of the subtracted $\HI$ intensity.
The \textcolor{black}{scattering is} fairly large with correlation coefficients of $\sim 0.6$ in the both plots (Figures~\ref{tau353whi1}(a) and \ref{tau353whi2}(a)).
Since $\WHI$ is supposed to be a good measure of the total $\HI$ in the optically thin approximation, $\taud$ should be highly correlated with $\WHI$ with a high correlation coefficient, if the $\HI$ is optically thin, and if gas and dust are well mixed with uniform properties.
}
The large scatter suggests that the dust properties may vary significantly, and{\slash}or that the optically thin approximation \textcolor{black}{for} the $\HI$ gas may not be valid.

\textcolor{black}{
In Figures~\ref{tau353whi1}(b), \ref{tau353whi1}(c), \ref{tau353whi2}(b), and \ref{tau353whi2}(c), we indicate $\Td$ for every $0.5\,\mathrm{K}$ interval at each point in Figures~\ref{tau353whi1}(a) and \ref{tau353whi2}(a).
}
We made least-squares fits between $\WHImain$ and $\taudmain$ by linear regression for each $\Td$ range.
Since the number of data points becomes less \textcolor{black}{for} the highest $\Td$, the fitting accuracy becomes worse for higher $\Td$.
In particular, for $\Td\geq 21.5\,\mathrm{K}$ the number of points becomes much less than for the others, and we here \textcolor{black}{assume} that the intercept is zero and \textcolor{black}{determine} only the slope.
The slopes $k$ ($\mathrm{K\,km\,s^{-1}}$), intercepts, and correlation coefficients are listed in Table 2.
We note the correlation coefficient is around $0.7\mbox{--}0.9$ for $\Td$ higher than $18\,\mathrm{K}$.
This clearly shows that the correlation strongly depends on $\Td$, and becomes better for the smaller $\Td$ ranges; generally speaking the slope becomes steeper and the intercept smaller with \textcolor{black}{increasing} $\Td$.

Figure~\ref{tau353tdwco}(a) shows a scatter plot of $\Td$ and $\taudmain$, and clearly indicates \textcolor{black}{a} significant decrease of $\Td$ with $\taudmain$.
A similar result that the dust temperature increases with decreasing gas column density has recently been found by the \textit{Planck} Collaboration \citep{2014A&A...571A..11P,2014A&A...566A..55P}.
Figure~\ref{tau353tdwco}(b) shows a scatter plot of $\WCO$ and $\taudmain$ in the area observed in $\CO$.
This shows positive correlation with a correlation coefficient of ${\sim}0.7$, while the dispersion is large.
The solid line in (b) represents a relationship $\FittaudWCO$, which is the result of a least-squares fit to the data with $\WCO\geq3\,\sigma$.

\section{Analysis}

From the behavior in Figure~\ref{tau353whi1} we infer that for the highest $\Td$ the optically thin approximation of $\HI$ is valid, whereas for lower $\Td$ the approximation becomes worse due to large $\HI$ optical depth.
Equation~(\ref{eq1}) is used to estimate $\NHI$ for the optically thin part in Figure~\ref{tau353whi1},
\begin{equation}
\NHI\,(\mathrm{cm^{-2}})=(\valXHI)\allowbreak\cdot\WHI\,(\mathrm{K\,km\,s^{-1}}),\label{eq1}
\end{equation}
and the relationship between $\taudmain$ and $\NHImain$ is estimated for the optically thin regime at $\Td\geq 21.5\,\mathrm{K}$;
\begin{equation}
\NHImain\,(\mathrm{cm^{-2}})=(\DerivedNHIcoef)\cdot\taudmain,\label{eq2}
\end{equation}
where $\DerivedNHIcoef$ is calculated by a product of $\valXHI$ in equation (\ref{eq1}) and the slope $k$ ($=8.47\times10^{7}\,\mathrm{K\,km\,s^{-1}}$) for $\Td>21.5\,\mathrm{K}$ in Table 2.

Relation (\ref{eq2}) holds as long as the dust properties are uniform, allowing us to calculate $\NHImain$ from $\taudmain$.
Then the coupled equations (\ref{eq3}) and (\ref{eq4}) in the following are used to solve \textcolor{black}{for} $\Ts$ and $\tauHImain$, where $\NHImain$ is calculated from $\taudmain$ by equation (\ref{eq2}),
\begin{eqnarray}
\WHImain\,(\mathrm{K\,km\,s^{-1}})&=&[\Ts\,(\mathrm{K})-\Tbg\,(\mathrm{K})]\allowbreak\cdot\dVHI\,(\mathrm{km\,s^{-1}}) \cdot\left[1-\rm{exp}(-\tauHImain)\right],\label{eq3}\\ \nonumber \\
\tauHImain&=&\frac{\NHImain\,(\mathrm{cm^{-2}})}{1.823\times 10^{18}}\cdot\frac{1}{\Ts\,(\mathrm{K})}\cdot\frac{1}{\dVHI\,(\mathrm{km\,s^{-1}})},\label{eq4}
\end{eqnarray}
where $\dVHI$, the $\HI$ linewidth, is given by $\WHImain/(\mathrm{peak\ \HI\ brightness\ temperature})$, and $\Tbg$ is the background continuum radiation temperature (see Section 2.4.) including the $2.7\,\mathrm{K}$ cosmic background radiation \citep{2009ApJ...707..916F}.
$\tauHImain$ is the $\HI$ optical depth averaged over the $\HI$ velocity width $\dVHI$, which ranges from $8\,\mathrm{km\,s^{-1}}$ to $30\,\mathrm{km\,s^{-1}}$ with a median at $\sim 14\,\mathrm{km\,s^{-1}}$.
Note that equation (\ref{eq4}) is valid not only for small $\tauHI$ but also for any positive value of $\tauHI$.
Figure~\ref{solutions} shows the two relationships in the $\Ts\mbox{-}\tauHImain$ plane for four typical values of $\Ts\simeq20\,\mathrm{K}$ (a), $40\,\mathrm{K}$ (b), $60\,\mathrm{K}$ (c), and $80\,\mathrm{K}$ (d).
We estimate the errors in $\Ts$ as shown in Figure~\ref{solutions} from the $1\,\sigma$ noise level of the $\HI$ and $\taudmain$ data.
We find that the equations give reasonable solutions with small errors for lower $\Ts$, whereas they cannot give a single set of $\Ts$ and $\tauHImain$ for $\Ts$ higher than $\sim 70\,\mathrm{K}$, where the error bars in $\Ts$ become uncomfortably large.
In Figure~\ref{solutions}, we estimate the error ranges as follows:
(a) $\Ts\simeq 20^{+4}_{-2}\,\mathrm{K}$, $\tauHImain\simeq 1.06^{+0.28}_{-0.29}$, 
(b) $\Ts\simeq 40^{+13}_{-6}\,\mathrm{K}$, $\tauHImain\simeq 0.84^{+0.28}_{-0.29}$,
(c) $\Ts\simeq 60^{+126}_{-20}\,\mathrm{K}$, $\tauHImain\simeq 0.31^{+0.21}_{-0.22}$, and
(d) $\Ts\simeq 80^{+\infty}_{-39}\,\mathrm{K}$, $\tauHImain\simeq 0.16^{+0.20}_{-0.16}$.
In the optically thin limit, the two equations become essentially the same, and can be satisfied by an infinite number of solutions, either a combination of large $\Ts$ and small $\tauHImain$, or that of small $\Ts$ and large $\tauHImain$.
Only the lower limit for $\Ts$ and the upper limit of $\tauHImain$ are constrained.

Figure~\ref{tstauhimap}(a) shows the distribution of $\Ts$, Figure~\ref{tstauhimap}(b) the distribution for $\tauHImain$, and Figure~\ref{tstauhimap}(c) the distribution of $\taudmain$.
Several specific areas are masked as described in Section 3 and also masked where the estimated $\Ts$ is greater than $70\,\mathrm{K}$.
In Figure~\ref{tstauhimap}(a) $\Ts$ is mostly lower than $50\,\mathrm{K}$.
We also find that $\tauHImain$ becomes large in many of the data points.
The results \textcolor{black}{for} $\Ts$ and $\tauHImain$ estimated are plotted in Figure~\ref{whinhi} with superposition of equation (\ref{eq3}) for $\Ts$ from $10\,\mathrm{K}$ to $100\,\mathrm{K}$ with $\tauHImain$.
We see saturation of $\WHImain$ becomes important for lower $\Ts$, and the slope for a given $\Ts$ decreases with \textcolor{black}{decreasing} $\Ts$ due to saturation.
This result is consistent with $\Ts$ estimated from comparisons between $\HI$ emission and absorption toward radio continuum sources, which show that cold $\HI$ clouds of $\TypicalTs$ are common \citep{2003ApJ...585..801D,2003ApJ...586.1067H}.
Their spatial coverage is however much smaller than the present coverage, $100\,\mathrm{square\mbox{-}degrees}$.
Possible hints of optically thick $\HI$ are also  found in the literature \citep{2004ApJ...603..560S,PlanetsstarsandstellarsystemsVolume5}.

\section{Physical conditions in the $\HI$ gas}

\subsection{$\Td$ and $\Ts$}

\textcolor{black}{The present analysis offers evidence for a significant fraction of cold $\HI$ gas in the region.}
The analysis shows the typical physical parameters of the optically thick $\HI$ gas as follows: 
$\NHImain=4\times10^{20}\,\mathrm{cm^{-2}}\mbox{--}1.5\times10^{21}\,\mathrm{cm^{-2}}$, 
$\HI$ number density $\nHI=\TypicalDensity$ for cloud line-of-sight depth of $3\,\mathrm{pc}$, and
$\Ts=\TypicalTs$.
The gas kinetic temperature of \textcolor{black}{the} $\HI$ is determined by heating due to the interstellar radiation field (ISRF) and cooling mainly by the \ion{C}{2} line \citep[e.g.,][]{1995ApJ...443..152W} and is in a range from $10\,\mathrm{K}\mbox{--}100\,\mathrm{K}$ for column density higher than $2\times 10^{20}\,\mathrm{cm^{-2}}$ \citep[e.g.,][]{2007ApJ...654..273G}.
Since the radiative transition rate, the Einstein $A$ coefficient, of the $21\,\mathrm{cm}$ line is small, $\sim 3\times 10^{-15}\,\mathrm{s^{-1}}$, the $21\,\mathrm{cm}$ transition is well thermalized by collisional excitation for density higher than \textcolor{black}{${\sim}10\,\mathrm{cm^{-3}}$} \citep[e.g.,][]{2001A&A...371..698L,1978AJ.....83.1607S}.
$\Ts$ is exactly in equilibrium with the gas kinetic temperature in the \textcolor{black}{dense} $\HI$ gas.
On the other hand, $\Td$ is determined by the balance between the ISRF heating and thermal radiation of dust grains and is generally in a range from $15\,\mathrm{K}$ to $25\,\mathrm{K}$ in the $\HI$ gas. $\Td$ and $\Ts$ should show correlation, since heating is commonly due to the ISRF, where $\Td$ variation is much smaller than that of $\Ts$ due to the strong temperature dependence of the dust cooling, as $(\Td)^{6}$ \citep*[e.g.,][]{Physicsoftheinterstellarandinergalaticmedium}.
For $\AV$ more than $1\,\mathrm{mag}$, $\Ts$ becomes less than $60\,\mathrm{K}$ \citep[e.g.,][]{2007ApJ...654..273G}, and $\Td$ shows significant decrease to $17\,\mathrm{K}$, which corresponds to $\Td$ within the $\CO$ clouds (Figure~\ref{maps}).
The present results on $\Ts$ and $\tauHImain$ are therefore consistent with the thermal properties of the ISM.
It is possible that $\Td$ is affected by the local radiation field in addition to the general ISRF.
The distribution of $\Td$ shown in Figure~\ref{maps}(c) indicates that in the northwest of the $\CO$ clouds $\Td$ tends to be higher than in the southeast.
%This may be due to the excess radiation field of the OB stars in the Sco OB2 association which are located $100\mbox{--}200\,\mathrm{pc}$ away from the present clouds \citep[c.f.,][]{2001AGM....18..P72T}.
This may be due to the excess radiation field of the OB stars in the Sco OB2 association which are located $100\mbox{--}200\,\mathrm{pc}$ away from the present clouds \citep[c.f.,][]{2001PASJ...53.1081T}.

An $\XCO$ factor converting $\WCO$ into $\NHtwo$ is estimated from a comparison of $\CO$ and $\taudmain$ (Figure~\ref{tau353tdwco}(b)).
A least-squares fit for $\WCO$ greater than $3.4\,\mathrm{K\,km\,s^{-1}}$ (above the $3\,\sigma$ noise level) yields $\FittaudWCO$.
\textcolor{black}{Here we adopt $\WCO$ greater than $3\,\sigma$ by considering that an $\XCO$ factor is conventionally estimated for $A_{V}\gtrsim2\mbox{--}3\,\mathrm{mag}$ \citep[e.g.][]{2006MNRAS.371.1865B}\textcolor{black}{.}
The offset $8.4\times10^{-6}$ in $\taudmain$ in the above relationship is interpolated as due to the contribution of the $\HI$, and the $\XCO$ factor is calculated from the slope $1.8\times10^{-6}$ in the relationship and the coefficient in equation (\ref{eq2}) as $\XCO=\NHtwo/\WCO=(1.5\times10^{26}/2)\cdot(1.8\times10^{-6})=\DerivedXCO\,\mathrm{cm^{-2}/(K\,km\,s^{-1})}$ with typical dispersion of $\sim\pm40\,\%$, which is somewhat smaller than that estimated elsewhere, $(2\mbox{--}3)\times10^{20}\,\mathrm{cm^{-2}/(K\,km\,s^{-1})}$ \citep[e.g.,][]{1993ApJ...416..587B}.}
Numerical simulations suggest that \textcolor{black}{the} $\XCO$ factor is relatively uniform in \textcolor{black}{regions} of high visual extinction where $\CO$ is intense \citep{2012ApJ...759...35I,2011MNRAS.412..337G,2006MNRAS.371.1865B}.
This difference may be possibly ascribed to the large contribution of cold $\HI$ gas which was not taken into account previously, and should be confirmed by a careful analysis of more sample clouds.
We note that we see a small number of points in Figure~\ref{tau353tdwco}(b) which show higher $\taudmain$ for a given $\WCO$ \textcolor{black}{on} the right of the dotted line and above the dashed line.
These may correspond to the molecular gas with slightly less $\CO$ abundance than the majority, suggesting a younger chemical evolutionary stage \citep{2003ApJ...592..217Y}, and can be characterized by a higher $\XCO$ factor.

\subsection{Alternative ideas}

The above interpretation assumes that all the neutral gas outside the $\CO$ is purely atomic.
This interpretation is consistent with the present analysis including only $\HI$ as shown by the fit of the line radiation transfer equation and estimated $\Ts$ and $\tauHImain$ (Figure~\ref{tstauhimap}).
We shall test if an interpretation based on $\CO$-free $\Htwo$ is possible as an alternative, which was advocated for a giant molecular cloud having total mass of $10^{5}\,M_{\odot}\mbox{--}10^{6}\,M_{\odot}$ by \citet{2010ApJ...716.1191W} and was discussed in the first \textit{Planck} paper \citep{2011A&A...536A..19P}.
$\taudmain$ is a sum of the dust opacity both in the $\HI$ and $\Htwo$ gas.
If $\CO$-free $\Htwo$ is dominant, $\Htwo$ may account for most of the $\taudmain$ instead of $\HI$.
The observed value of $\WHImain$ poses a strict lower limit on $\NHImain$ and it is not probable that $\Htwo$ dominates $\HI$.
The majority of the data points have $\Ts\sim 30\,\mathrm{K}$ (Figure~\ref{whinhi}) and a lower limit for their $\HI$ column density is given by the optically thin limit.
For instance, for points having $\WHImain=250\,\mathrm{K\,km\,s^{-1}}$, $\NHImain$ at $\Ts\sim 30\,\mathrm{K}$ is $0.85\times 10^{21}\,\,\mathrm{cm^{-2}}$, whereas $\NHImain$ in the optically thin limit is $0.45\times 10^{21}\,\mathrm{cm^{-2}}$ corresponding to ${\sim}50\,\%$ of $0.85\times 10^{21}\,\,\mathrm{cm^{-2}}$, and for points having $\WHImain=200\,\mathrm{K\,km\,s^{-1}}$, $\NHImain$ at $\Ts\sim 30\,\mathrm{K}$ is $0.6\times 10^{21}\,\mathrm{cm^{-2}}$ whereas $\NHImain$ in the optically thin limit is $0.35\times 10^{21}\,\mathrm{cm^{-2}}$ corresponding to ${\sim}60\,\%$ of $0.6\times 10^{21}\,\,\mathrm{cm^{-2}}$.
This suggests that the cold gas typically has $\Htwo$ abundance of $50\,\%$ or less over most of the region outside of the $\CO$ clouds.
Direct UV absorption measurements of $\Htwo$ abundances on the line of sight toward NGC\,7469 ($\ell=83\fdg10, b=-45\fdg47$) show $\NHtwo=4.7\times10^{19}\,\mathrm{cm^{-2}}$ and $\NHI=3.9\times10^{20}\,\mathrm{cm^{-2}}$ \citep{2006ApJ...636..891G}, indicating that the hydrogen is about $90\,\%$ atomic.
It is worth noting that other lines of sight at intermediate latitudes observed by \textit{FUSE} and \textit{Copernicus} \citep[summarized by][]{2009ApJS..180..125R} show similarly low molecular fractions unless the total H column density, $\NHtwo+\NHI\ga10^{21}\,\mathrm{cm^{-2}}$, lending support for the conclusion that $\HI$ is dominant.
Finally, we note that the timescale for $\Htwo$ formation is generally as long as $10^{7}\,\mathrm{years}$ for typical density $100\,\mathrm{cm^{-3}}$ \citep{1971ApJ...163..165H}.
The cloud lifetime is however smaller than $\sim10^{6}\,\mathrm{years}$ as estimated from the crossing time at latitude higher than $10\,\mathrm{degrees}$ \textcolor{black}{for} a typical cloud size of several $\mathrm{pc}$ and a line width of $15\,\mathrm{km\,s^{-1}}$ \citep{2003ApJ...592..217Y,2006ApJ...642..307Y}.
This suggests that the present cloud is too young to form significant $\Htwo$. 
Another possibility is that the dust properties may be considerably different in the region, \textcolor{black}{as} has been explored by the \textit{Planck} collaboration \citep{2014A&A...571A..11P}.
We shall defer \textcolor{black}{exploration of} this possibility until a full account of the study is opened to the community.

\subsection{Mass of the $\HI$ envelope}

We estimate the total mass of the clouds in the region under study.
We estimate the total system mass including $\HI$, $\Htwo$, and $\mathrm{He}$ ($40\,\%$ of hydrogen in mass) to be $\Mtotal$, by using the sum of $\taudmain$ above $5\times 10^{-6}$ under equation (\ref{eq2}) applied both to $\HI$ and $\Htwo$ (Figure~\ref{tstauhimap}(c)).
The typical ratio of the opacity-corrected case to the optically thin case is estimated to be $\sim2.7$ toward areas without $\CO$ emission, showing that the opacity correction has a significant impact.
The mass of the $\HI$ gas in the masked areas is interpolated to be $\sim9.0\times 10^{2}\,M_{\odot}$ by using the average $\NHImain\sim1.3\times 10^{21}\,\mathrm{cm^{-2}}$ in the areas without masking, and corresponds to $\sim7\,\%$ of the total system mass above.
On the other hand, the molecular mass is estimated to be $\MHtwo$ by applying the present $\XCO$ factor ($=\DerivedXCO\,\mathrm{cm^{-2}/(K\,km\,s^{-1})}$) to the $\CO$ emission.
\textcolor{black}{Here we integrated the $\CO$ emission above the $3\,\sigma$ noise fluctuations of $3.4\,\mathrm{K\,km\,s^{-1}}$ in the integrated intensity in the $5\,\mathrm{arcmin}$ smoothed beam.}
By subtracting this mass from the total system mass, we obtain $\MHIenv$ as the total mass of the $\HI$ envelope.
We thus find that the $\HI$ envelope has a total mass $\sim10$ times larger than the mass of the $\Htwo$ clouds probed by $\CO$.
\textcolor{black}{
Note that in \citet{1994ApJ...434..162G} the mass of the $\HI$ envelope was calculated as $\sim3\times 10^{3}\,M_{\odot}$, which is smaller than the present result.
This difference may be attributed to the fact that (i) the area they used for the calculation corresponds to ${\sim}1/2$ of that used in our calculation, and (ii) they assumed that all the $\HI$ gas is optically thin.
}

We conclude that the $\Htwo$ clouds traced by $\CO$ are surrounded by massive, optically thick $\HI$ envelopes possibly containing a relatively small fraction of $\Htwo$ that has no corresponding $\CO$ emission \textcolor{black}{above the $3\,\sigma$ noise fluctuations of $3.4\,\mathrm{K\,km\,s^{-1}}$}.
Figure~\ref{histograms} is a histogram of the mass of the $\HI$ envelope as a function of $\Ts$.
This clearly shows that most of the $\HI$ \textcolor{black}{is in} the $\Ts$ ranges from $20\,\mathrm{K}$ to $40\,\mathrm{K}$.
Assuming the depth of the $\HI$ envelope along the line of sight to be $3\,\mathrm{pc}$, the $\HI$ density is estimated to be $\TypicalDensity$.
This density is consistent with $\Ts=\TypicalTs$ in a model calculation of the thermal balance \citep{2007ApJ...654..273G}.

It is generally known that small clouds having mass less than $\sim1000\,M_{\odot}$ cannot be gravitationally bound due to the large virial mass as compared with the $\CO$ luminosity mass \citep[e.g.,][]{1997ApJS..110...21Y,1998ApJS..117..387K}.
The $\HI$ envelope may have a deep influence on cloud dynamics.
The average density of the $\HI$ is $100\,\mathrm{cm^{-3}}$ and the velocity width is $15\,\mathrm{km\,s^{-1}}$.
The dynamical pressure of $\HI$, $\rho\,v^{2}$ $[(\mathrm{density})\cdot(\mathrm{velocity\ dispersion})^{2}]$, is nearly equal to that of the $\Htwo$ gas and the pressure exerted by the $\HI$ envelope is able to confine the $\Htwo$.
The $\HI$ envelope may \textcolor{black}{help} to stabilize small $\Htwo$ clouds.
It is yet perhaps true that the whole system including the $\HI$ envelope is not gravitationally bound because of the large velocity dispersion of the $\HI$, and some dynamically transient states, like those in colliding $\HI$ flows, must be invoked to explain the cloud dynamics \citep[e.g.,][]{2001ApJ...562..852H}.

\subsection{Relation to the dark gas}

Recent analyses of gamma ray maps and the \textit{Planck} data show \textcolor{black}{evidence for} dark gas which is not seen in $\CO$ or $\HI$ emission \citep{2005Sci...307.1292G,2011A&A...536A..19P}.
These analyses did not consider the $\HI$ opacity effect and assumed that $\HI$ is completely optically thin.
The present study has shown that the $\HI$ optical depth effect is important in \textcolor{black}{this} region and an upward correction of the cloud mass by a factor of $\sim2$ on average is required.
A significant fraction of the relatively dense $\HI$ gas may therefore be masked by saturation of the $21\,\mathrm{cm}$ line, and the cold $\HI$ gas is a viable candidate for the dark gas.
This possibility deserves further exploration by including a much larger fraction of the sky.
We note that the present analysis is applicable in a $\taudmain$ range from $2\times 10^{-6}$ to $\mathrm{a\ few \times 10^{-5}}$.
Therefore, the behavior of the $\HI$ outside this range is yet to be explored separately.

\section{Conclusions} 

We have made a comparison of $\HI$, $\CO$, and the \textit{Planck} dust properties in a complex of high latitude clouds \object{MBM 53, 54, 55} and \object{HLCG $92{-}35$} and conclude from this study as follows;

The $\HI$ intensity shows a poor correlation with dust opacity with a correlation coefficient of $\sim 0.6$.
We interpret that this is caused by the effect of the optical depth on $21\,\mathrm{cm}$ $\HI$ emission.
We present a method to estimate the spin temperature and the optical depth of the $\HI$ emission which couples the equation of line radiation transfer and the expression for the $\HI$ optical depth by assuming that dust properties do not change significantly in the region, where the $\taudmain$ is limited in a range from $3\times 10^{-6}$ to $2\times 10^{-5}$.
We have analyzed about $8\times 10^{5}$ data points smoothed at $5\,\mathrm{arcmin}$ resolution and successfully estimated $\Ts$ and $\tauHImain$.
Most of the points have $\Ts$ in \textcolor{black}{the} range from $20\,\mathrm{K}$ to $40\,\mathrm{K}$ and $\tauHImain$ is in \textcolor{black}{the} range from $0.2$ to $4$.
This indicates that saturation due to large $\HI$ optical depth provides a reasonable explanation of the suppressed $\HI$ intensity in regions with large dust opacity.
The $\Ts$ distribution is consistent with previous results estimated from absorption measurements toward radio continuum sources while \textcolor{black}{in this paper the spatial coverage} is continuous, covering a much larger area than \textcolor{black}{continuum source} absorption measurements.
If the present results are correct, this suggests that optically thick $\HI$ is more significant than previously assumed in the literature.
The typical physical parameters of the cold $\HI$ are density $\TypicalDensity$, $\Ts=\TypicalTs$ and $\NHImain=4\times10^{20}\,\mathrm{cm^{-2}}\mbox{--}1.5\times10^{21}\,\mathrm{cm^{-2}}$.
The $\XCO$ factor to covert $\WCO$ into $\NHtwo$ has been estimated to be $\DerivedXCO\,\mathrm{cm^{-2}/(K\,km\,s^{-1})}$.
The $\CO$ clouds are associated with massive dense cold $\HI$ envelope having $\MHIenv$ as compared with the $\CO$ cloud mass of $\MHtwo$.
Another possibility, that the dust properties may vary significantly, is \textcolor{black}{being} explored currently by the \textit{Planck} collaboration, and should be considered in depth as a future step.
Finally, it is an obvious task to extend this method to the whole sky to see if cold $\HI$ is common and dominant in the local interstellar space around the Sun.

\acknowledgements

We are grateful to Fran\c{c}ois Boulanger and Jean-Phillippe Bernard for their initiative to begin collaboration between the \textit{Planck} and NANTEN2.
The high latitude clouds in the paper were chosen as one of the primary targets in the comparative study. 
We thank the anonymous referee for the valuable comments on the paper.

NANTEN2 is an international collaboration among 10 universities; Nagoya University, Osaka Prefecture University, University of Cologne, University of Bonn, Seoul National University, University of Chile, University of New South Wales, Macquarie University, University of Sydney, and University of ETH Zurich.
This work was financially supported by Grants-in-Aid for Scientific Research (KAKENHI) of Japanese society for the Promotion of Science (JSPS) (grant numbers 24224005, 25287035, 23403001, 23540277, and 23740149-01).
This work was also financially supported by the Young Research Overseas Visits Program for Vitalizing Brain Circulation (R2211) and the Institutional Program for Young Researcher Overseas Visits (R29) by the Japan Society for the Promotion of Science (JSPS) and by the grant-in-aid for Nagoya University Global COE Program, ``Quest for Fundamental Principles in the Universe: From Particles to the Solar System and the Cosmos,'' from MEXT.
Based on observations obtained with \textit{Planck} (\url{http://www.esa.int/Planck}), an ESA science mission with instruments and contributions directly funded by ESA Member States, NASA, and Canada.
This publication utilizes data from Galactic ALFA $\HI$ (GALFA $\HI$) survey data set obtained with the Arecibo L-band Feed Array (ALFA) on the Arecibo $305\,\mathrm{m}$ telescope. Arecibo Observatory is part of the National Astronomy and Ionosphere Center, which is operated by Cornell University under Cooperative Agreement with the U.S. National Science Foundation.
The GALFA $\HI$ surveys are funded by the NSF through grants to Columbia University, the University of Wisconsin, and the University of California.
Some of the results in this paper have been estimated using the {\sc healpix} \citep{2005ApJ...622..759G} package.

\section*{Appendix A}

Figure~\ref{chmaps} shows five $\HI$ velocity channel distributions from $V_{\mathrm{LSR}}=-45\,\mathrm{km\,s^{-1}}$ to $+32\,\mathrm{km\,s^{-1}}$.
The main $\HI$ cloud toward the $\CO$ emitting clouds is seen from $-12.7\,\mathrm{km\,s^{-1}}$ to $-0.3\,\mathrm{km\,s^{-1}}$ (panel (b)) as noted by \citet{1994ApJ...434..162G}.
The main feature is outstanding in the $\HI$ intensity having $200\,\mathrm{K\,km\,s^{-1}}$ in the integrated intensity.
In this velocity channel, we see additional $\HI$ features; one is located $\sim5\,\mathrm{degree}$ west of the main cloud and is elongated in a similar direction and with similar length to the main cloud (called West1 $\HI$), and the other in the north of the main cloud peaked at $(l, b)\sim(100^{\circ}, -31^{\circ})$ (North $\HI$). West1 $\HI$ seems to be linked with the northern tip of the $\CO$ clouds.
We also find an additional feature (West2 $\HI$) peaked at $(l, b)\sim(90^{\circ}, -37^{\circ})$ which may be extended to the west up to $l\sim80^{\circ}$. 

In $V_{\mathrm{LSR}}$ from $-25.0\,\mathrm{km\,s^{-1}}$ to $-12.7\,\mathrm{km\,s^{-1}}$ (panel (a)), we see two features; one of them corresponds to North $\HI$ in position and the other to West2 $\HI$.
In addition, the $\HI$, North $\HI$ and several smaller peaks along the southern rim of the main cloud, seems to be surrounding the $\CO$ as noted by \citet{2003ApJ...592..217Y} (see also their Figure 6). 

\textcolor{black}{
In the velocity from $-0.3\,\mathrm{km\,s^{-1}}$ to $+12.0\,\mathrm{km\,s^{-1}}$ (panel (c)), we see the $\HI$ corresponds to West1 $\HI$ and the $\HI$ in the north shows intensity depression surrounding the $\CO$, suggesting physical association with the $\CO$. 
}

In the other extreme velocity range from $-45.0\,\mathrm{km\,s^{-1}}$ to $-25.0\,\mathrm{km\,s^{-1}}$ (panel (d)) we find the main peak at $(l, b)\sim(87^{\circ}, -38^{\circ})$ shows elongation similar to West2 $\HI$.
The $\HI$ may be possibly linked with West2, while it was not included in the present analysis. 

We summarize the above that the $\HI$ in the three panels (a), (b) and (c) is likely associated with the main cloud within a volume having a size in the order of the main $\HI$ cloud, $30\mbox{--}40\,\mathrm{pc}$, at a distance of $150\,\mathrm{pc}$. 

\section*{Appendix B}
\textcolor{black}{
$\taudmain$, which is the main component of $\taud$, is calculated as follows;
\begin{itemize}
\item[(1)] First we use slope derived by least-squares fits between $\WHI$ and $\taud$ by linear regression
for $\Td > 21.5\,\mathrm{K}$ in Figure~\ref{tau353whi2}(b).
\item[(2)] By using the slope derived from (1) $\taud^{\prime}$ is calculated by $\taud^{\prime}=\taud-\WHI\,(V_{\mathrm{LSR}}<\HIVoneround\,\mathrm{km\,s^{-1}}, \HIVtworound\,\mathrm{km\,s^{-1}}<V_{\mathrm{LSR}})/(\mathrm{slope})$.
\item[(3)] We estimate new slope again by the same manner as (1) between $\WHImain$ and $\taud^{\prime}$.
\item[(4)] $\taud^{\prime\prime}$ is calculated by the same manner as (2) but by using $\taud^{\prime}$ and the slope derived by (3).
\item[(5)] Iterating (3) and (4) until the value of the slope becomes converged.
\end{itemize}
After 5 iterations the best estimate of the slope $k$ converged to $8.34\times10^{7}\,\mathrm{K\,km\,s^{-1}}$ with the relative variation less than $10^{-4}$.
}

\bibliographystyle{apj} \bibliography{reference}

\clearpage

\begin{figure}
\epsscale{0.4}
\plotone{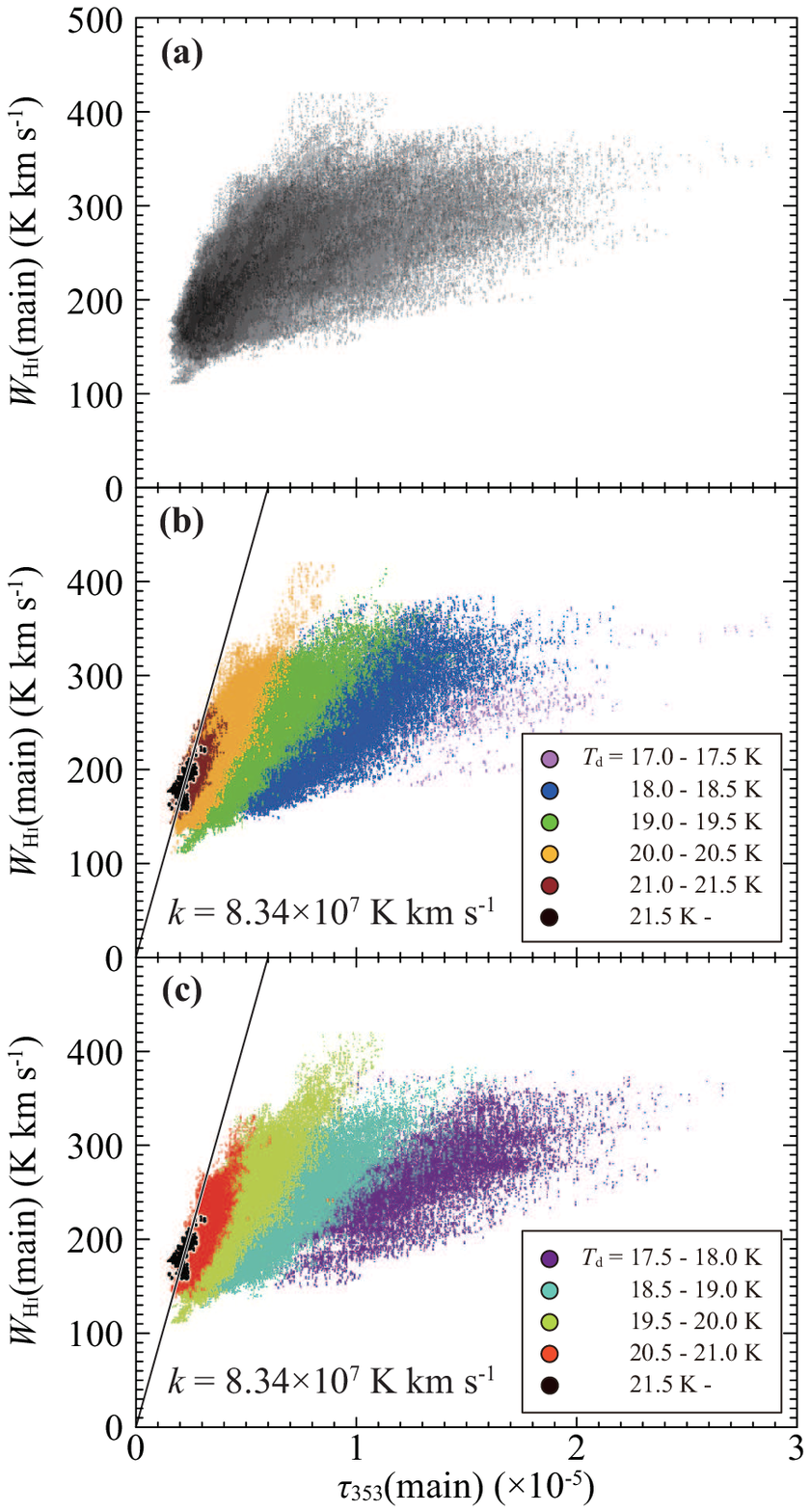}
\caption{Scatter plots between $\WHImain$ and $\taudmain$ and between $\WHI$ and $\taud$ in the region shown in Figure~\ref{maps}. Specific areas are masked. Details on the masking are described in \textcolor{black}{Section 2.5}. {\bf Figure~\ref{tau353whi1}:} scatter plots of $\WHImain$ whose integrated velocity range is from $\HIVoneround\,\mathrm{km\,s^{-1}}$ to $\HIVtworound\,\mathrm{km\,s^{-1}}$ and $\taudmain$, which is the $\taud$ component associated with the $\HI$ whose velocity range is from $\HIVoneround\,\mathrm{km\,s^{-1}}$ to $\HIVtworound\,\mathrm{km\,s^{-1}}$. Details on how to calculate this $\taudmain$ value are described in the Appendix B. {\bf (a)} plots in black dots for all points. {\bf (b)} and {\bf (c)} show the scatter plots for $\Td$ in windows of $0.5\,\mathrm{K}$ interval every $1\,\mathrm{K}$.
In order to avoid heavy overlapping among the data points, in {\bf (b)} we show five $\Td$ ranges, $17.0\,\mathrm{K}\mbox{--}17.5\,\mathrm{K}$ (light purple), $18.0\,\mathrm{K}\mbox{--}18.5\,\mathrm{K}$ (blue), $19.0\,\mathrm{K}\mbox{--}19.5\,\mathrm{K}$ (green), $20.0\,\mathrm{K}\mbox{--}20.5\,\mathrm{K}$ (yellow), and $21.0\,\mathrm{K}\mbox{--}21.5\,\mathrm{K}$ (brown), and in {\bf (c)} five $\Td$ ranges shifted by $0.5\,\mathrm{K}$, $17.5\,\mathrm{K}\mbox{--}18.0\,\mathrm{K}$ (purple), $18.5\,\mathrm{K}\mbox{--}19.0\,\mathrm{K}$ (light blue), $19.5\,\mathrm{K}\mbox{--}20.0\,\mathrm{K}$ (light green), and $20.5\,\mathrm{K}\mbox{--}21.0\,\mathrm{K}$ (red).
In the both panels {\bf (b)} and {\bf (c)} the $\Td$ range, $21.5\,\mathrm{K}$ and higher, is shown in black along with the linear regression line obtained by the least-squares fit which is assumed to have zero intercept.
{\bf Figure~\ref{tau353whi2}:} same as Figure~\ref{tau353whi1} but integrated velocity range of the $\HI$
is from $\MinHIVR\,\mathrm{km\,s^{-1}}$ to $\MaxHIVR\,\mathrm{km\,s^{-1}}$ and $\taud$ is the total value.}
\label{tau353whi1}
\end{figure}

\begin{figure}
\epsscale{0.4}
\plotone{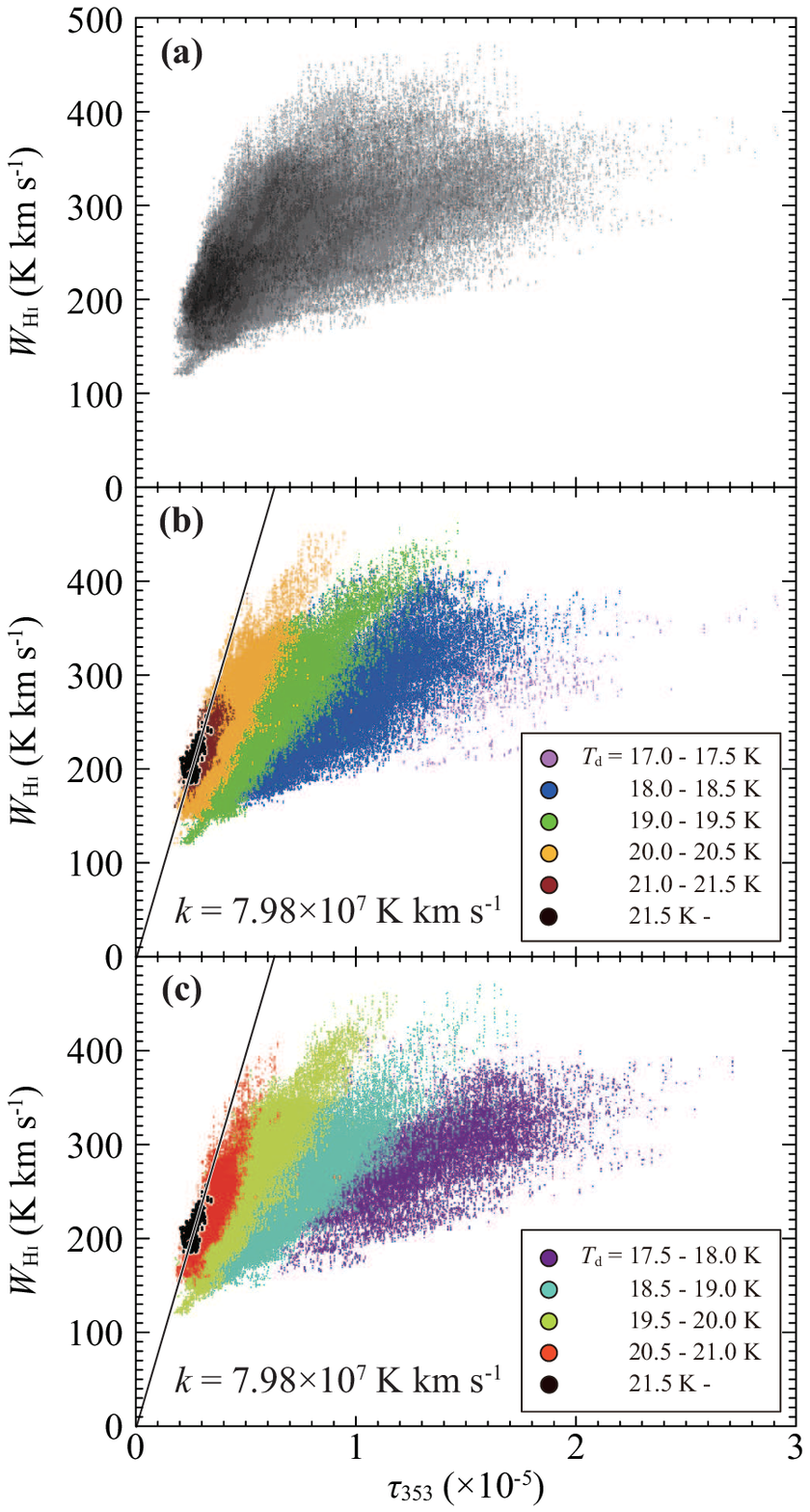}
%\caption{(This figure should be placed next to Figure~\ref{tau353whi1})}
\caption{}
\label{tau353whi2}
\end{figure}

\begin{figure}
\epsscale{0.8}
\plotone{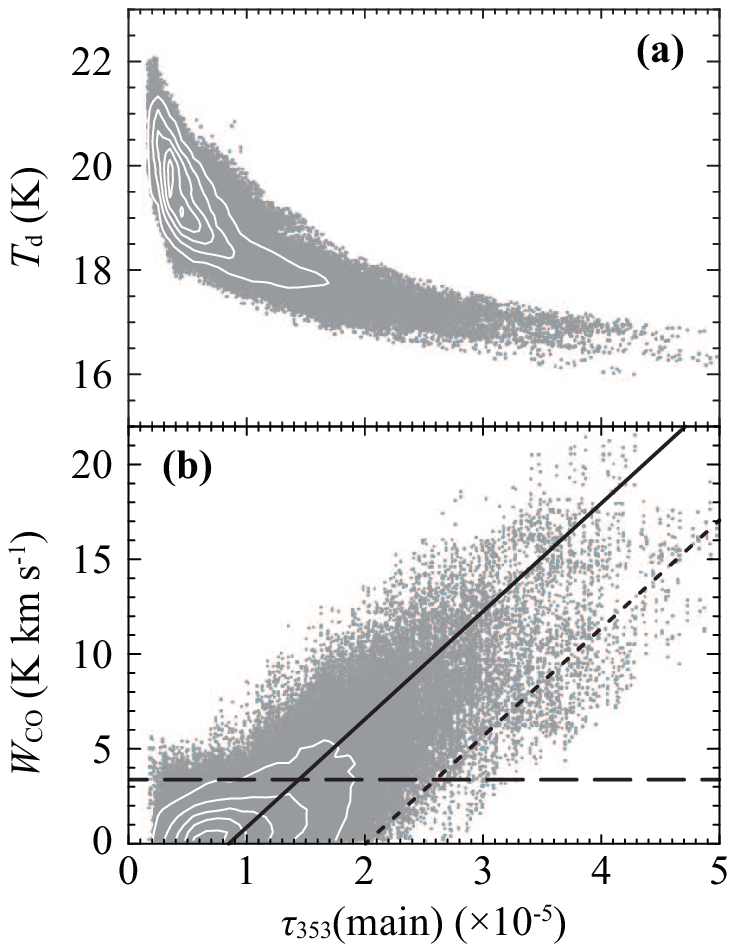}
\caption{\textcolor{black}{{\bf (a)} Correlation plot between $\Td$ and $\taudmain$ in the area shown in Figure~\ref{tstauhimap}(c). The contours represent the $90\,\%$, $70\,\%$, $50\,\%$, $30\,\%$, and $10\,\%$ levels of the maximum of the density of the plotted data over the ($\Td$, $\taudmain$) space. {\bf (b)} same as {\bf (a)} but for $\WCO$ and $\taudmain$. The contours represent the $90\,\%$, $70\,\%$, $50\,\%$, $30\,\%$, and $10\,\%$ levels of the maximum of the density of the plotted data over the ($\WCO$, $\taudmain$) space. The dashed line represents $3\,\sigma$ level of the $\CO$ data ($3.4\,\mathrm{K\,km\,s^{-1}}$), and the solid line a relationship $\FittaudWCO$, which is the result of a least-squares fit to the data with $\WCO>3\,\sigma$. The dotted line is $\taudmain=(1.8\times10^{-6})\cdot\WCO\,(\mathrm{K\,km\,s^{-1}})+(2\times10^{-5})$ (see text).}}
\label{tau353tdwco}
\end{figure}

\begin{figure}
\epsscale{1.0}
\plotone{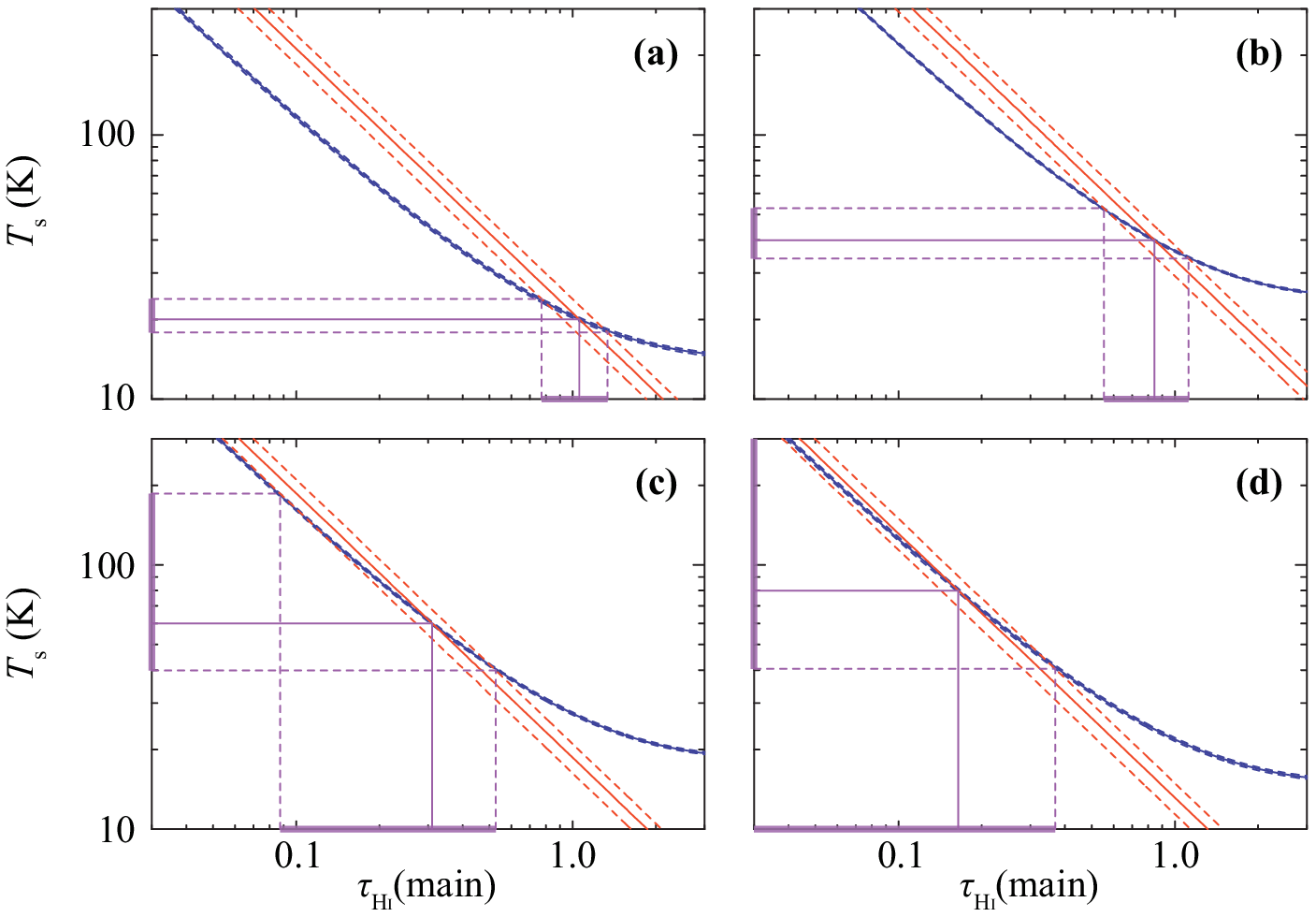}
\caption{\textcolor{black}{The blue and red lines represent equations {\bf (3)} and {\bf (4)} in the text, respectively. The crossing point of the two lines is the solution that satisfies both equations {\bf (3)} and {\bf (4)}. {\bf (a)}, {\bf (b)}, {\bf (c)}, and {\bf (d)} show the typical cases of the solutions of $\Ts\simeq 20\,\mathrm{K}$, $\simeq 40\,\mathrm{K}$, $\simeq 60\,\mathrm{K}$, and $\simeq 80\,\mathrm{K}$, and are located at $(\ell, b)\sim(92\fdg70, -41\fdg56)$, $(90\fdg07, -42\fdg68)$, $(89\fdg69, -29\fdg12)$, and $(86\fdg03, -30\fdg09)$, respectively. The dashed lines around each line are the error taking observational parameters into consideration. The purple solid lines indicate the solutions of $\tauHImain$ and $\Ts$, and dashed lines their errors.}}
\label{solutions}
\end{figure}

\begin{figure}
\epsscale{0.5}
\plotone{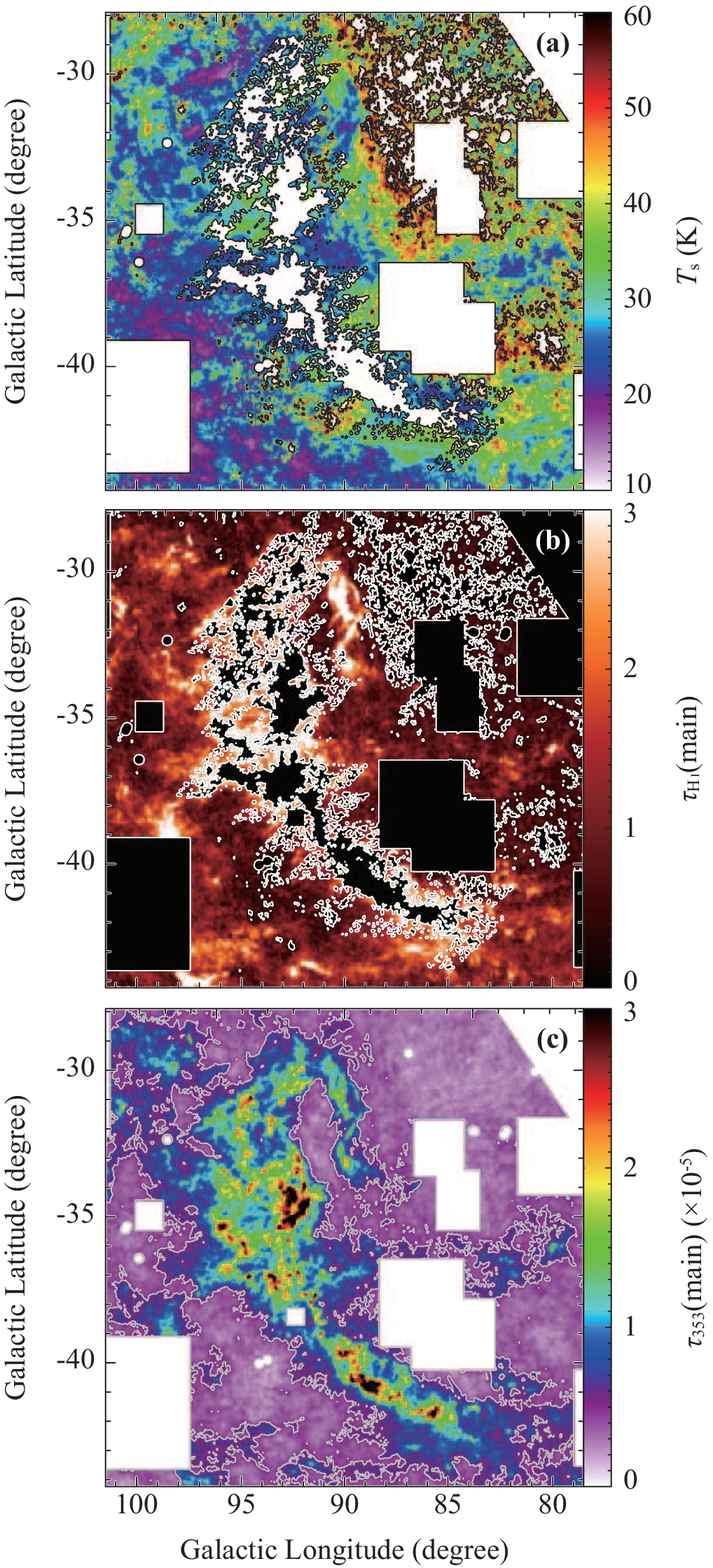}
\caption{\textcolor{black}{{\bf (a)} and {\bf (b)} show the spatial distribution of $\Ts$ and $\tauHImain$ solved from the two equations (3) and (4) in the text. The masks described in Section 2.5. are applied, and in addition the areas where $\Ts\geq70\,\mathrm{K}$ are masked. {\bf (c)} represents the spatial distribution of $\taudmain$. The gray contours indicate the level of $\taudmain=5\times10^{-6}$.}}
\label{tstauhimap}
\end{figure}

\begin{figure}
\epsscale{1.0}
\plotone{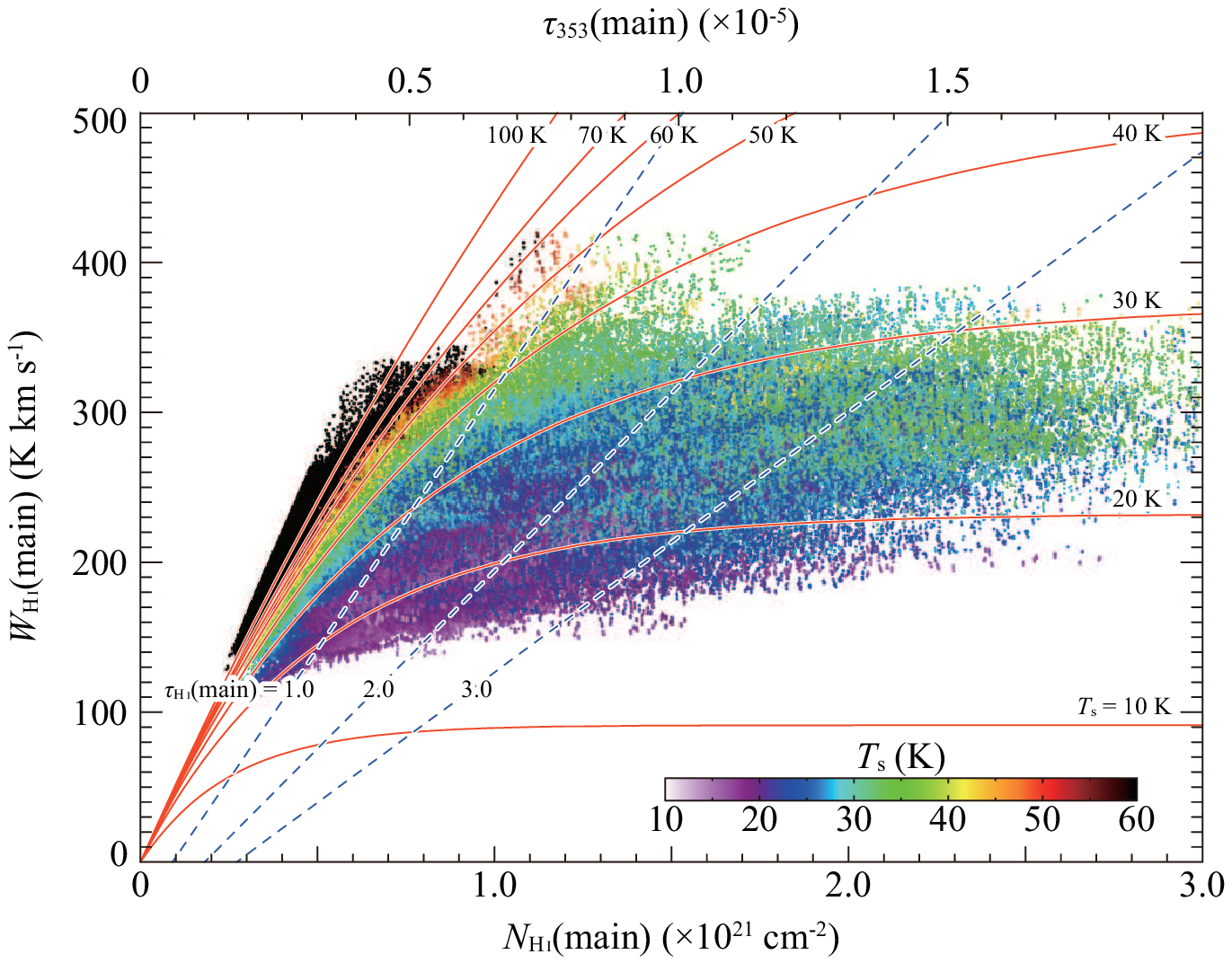}
\caption{\textcolor{black}{Correlation plot between $\WHImain$ and $\NHImain$. The data in the region shown in the Figure~\ref{tstauhimap}(a)(b) is plotted. The color of each point represents $\Ts$. The solid and dashed lines represent $\Ts$ and $\tauHImain$ calculated from equations (3) and (4) in the text by assuming $\dVHI$ of $\sim14.1\,\mathrm{km\,s^{-1}}$ (median value) and $\Tbg$ of $\sim3.5\,\mathrm{K}$ (median value), respectively.}}
\label{whinhi}
\end{figure}

\begin{figure}
\epsscale{1.0}
\plotone{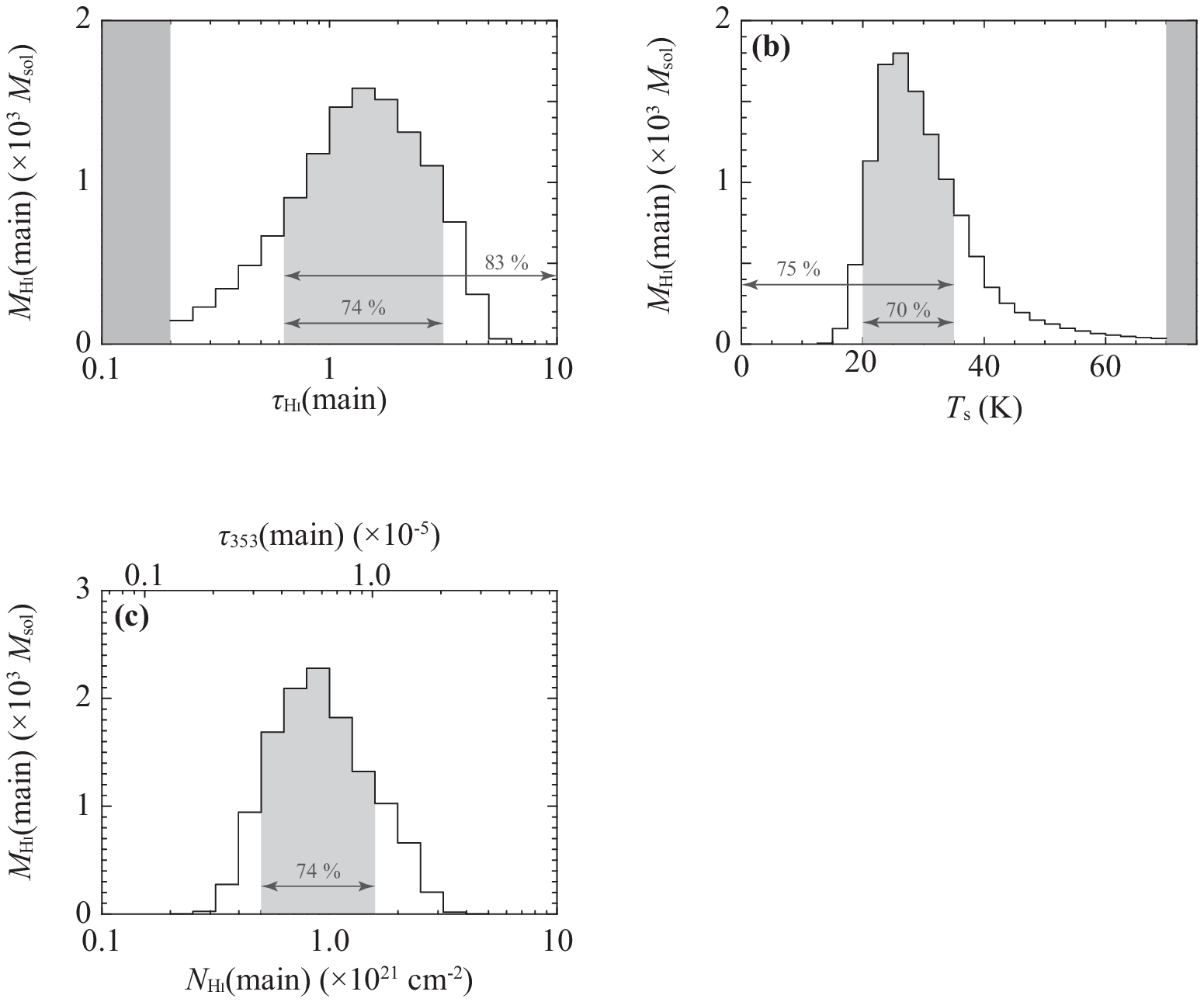}
\caption{\textcolor{black}{{\bf (a)} Histogram of the mass of the $\HI$ envelope as a function of $\tauHImain$. {\bf (b)} The same as {\bf (a)}, but as a function of $\Ts$. {\bf (c)} The same as (a) and (b), but as functions of $\NHImain$ and $\taudmain$. The areas filled with light-gray are determined by the widths at the half-levels of the each histogram. Note that in panels (a) and (b), the histograms where $\tauHImain\leq0.2$ or $\Ts\geq70\,\mathrm{K}$ are masked because the solutions of $\tauHImain$ and $\Ts$ are not well-estimated.}}
\label{histograms}
\end{figure}

\begin{figure}
\epsscale{1.0}
\plotone{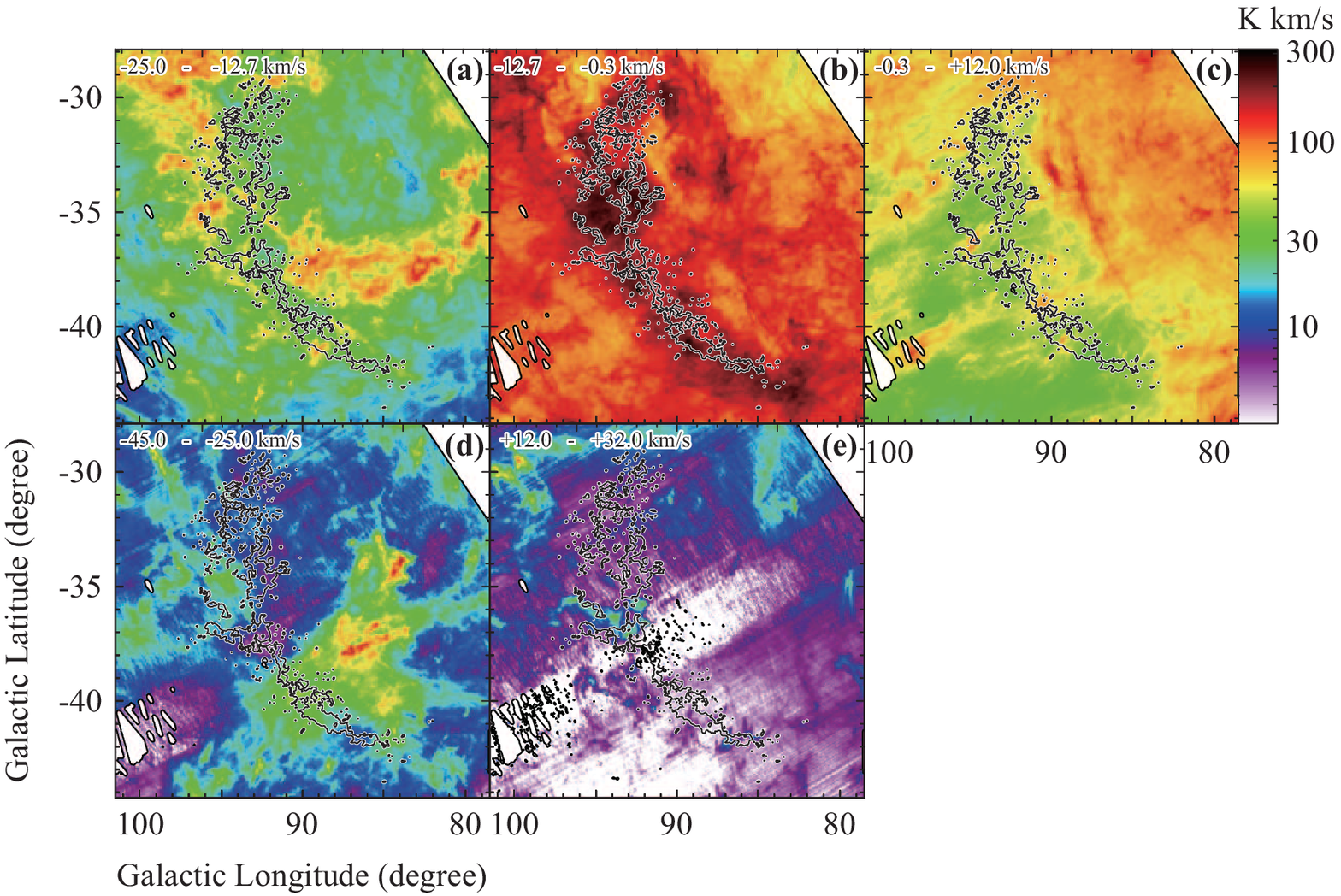}
\caption{\textcolor{black}{Velocity channel maps of the $\HI$ (color image) superposed on the $\WCO$ (contours). In panels (a), (b), and (c), we show the channel maps of the main $\HI$ cloud with dividing the velocity range into three. Panels (d) and (e) show the channel maps outside the main $\HI$ velocity range. The contour level of $\WCO$ is $3.4\,\mathrm{K\,km\,s^{-1}}$ for each panel.}}
\label{chmaps}
\end{figure}

%% The values (usually only l,r and c) in the last part of
%% \begin{deluxetable}{} command tell LaTeX how many columns
%% there are and how to align them.
\begin{deluxetable}{ccccc}
%% Keep a portrait orientation
%% Over-ride the default font size
%% Use Default (12pt)
%% Use \tablewidth{?pt} to over-ride the default table width.
\tablewidth{0pt}
%% If you are unhappy with the default look at the end of the
%% *.log file to see what the default was set at before adjusting
%% this value.
%% This is the title of the table.
\tablecaption{List of masked areas\label{table1}}
%% This command over-rides LaTeX's natural table count
%% and replaces it with this number.  LaTeX will increment 
%% all other tables after this table based on this number
\tablenum{1}
%% The \tablehead gives provides the column headers.  It
%% is currently set up so that the column labels are on the
%% top line and the units surrounded by ()s are in the 
%% bottom line.  You may add more header information by writing
%% another line between these lines. For each column that requries
%% extra information be sure to include a \colhead{text} command
%% and remember to end any extra lines with \\ and include the 
%% correct number of &s.
\tablehead{\multicolumn{2}{c}{position} & \colhead{Area} & \colhead{\phm{mm}Object name\phm{mm}} & \colhead{\phm{mmm}Remark\phm{mmm}} \\ 
\colhead{$\ell\,(\mathrm{degree})$} & \colhead{$b\,(\mathrm{degree})$} & \colhead{$(\mathrm{degree^{2}})$} & \colhead{} & \colhead{}}
%% All data must appear between the \startdata and \enddata commands
\startdata
$79.1\phn$ & $-41.8\phn$ & $\phn1.1$ & \multicolumn{1}{c}{\nodata} & $(a)$ \\
$81.5\phn$ & $-29.3\phn$ & $\phn6.5$ & \multicolumn{1}{c}{\nodata} & $(a)$ \\
$81.5\phn$ & $-33.0\phn$ & $\phn6.0$ & \multicolumn{1}{c}{\nodata} & $(b)$ \\
$85.6\phn$ & $-33.8\phn$ & $\phn6.4$ & \multicolumn{1}{c}{\nodata} & $(b)$ \\
$86.1\phn$ & $-38.2\phn$ & $13.1$ & \multicolumn{1}{c}{3C 454.3} & $(c)$ \\
$92.2\phn$ & $-38.9\phn$ & $\phn0.3$ & \multicolumn{1}{c}{\nodata} & $(a)$ \\
$98.3\phn$ & $-35.1\phn$ & $\phn1.0$ & \multicolumn{1}{c}{\nodata} & $(a)$ \\
$99.1\phn$ & $-41.6\phn$ & $13.1$ & \multicolumn{1}{c}{\nodata} & $(a)$ \\
$99.4\phn$ & $-30.1\phn$ & $\phn0.7$ & \multicolumn{1}{c}{\nodata} & $(a)$ \\ \hline
$80.58$ & $-32.47$ & $8.5\times10^{-2}$ & \multicolumn{1}{c}{IRAS 22221+1748} & infrared source \\
$82.42$ & $-30.16$ & $8.5\times10^{-2}$ & \multicolumn{1}{c}{33 Peg} & double or multiple star \\
$83.37$ & $-32.25$ & $8.6\times10^{-2}$ & \multicolumn{1}{c}{Mrk 306} & galaxy \\
$83.46$ & $-32.39$ & $8.6\times10^{-2}$ & \multicolumn{1}{c}{HD 213618} & star \\
$84.67$ & $-32.34$ & $8.7\times10^{-2}$ & \multicolumn{1}{c}{HD 214128} & star \\
$84.73$ & $-32.31$ & $8.7\times10^{-2}$ & \multicolumn{1}{c}{NGC 7316} & galaxy \\
$86.05$ & $-33.18$ & $8.6\times10^{-2}$ & \multicolumn{1}{c}{Mrk 308} & Seyfert galaxy \\
$86.09$ & $-33.57$ & $8.6\times10^{-2}$ & \multicolumn{1}{c}{BD+19 4992} & star \\
$87.46$ & $-29.73$ & $8.6\times10^{-2}$ & \multicolumn{1}{c}{NGC 7339} & radio galaxy$^{(d)}$ \\
$93.53$ & $-40.35$ & $8.7\times10^{-2}$ & \multicolumn{1}{c}{RAFGL 3068} & variable star \\
$93.91$ & $-40.47$ & $8.7\times10^{-2}$ & \multicolumn{1}{c}{NGC 7625} & interacting galaxies \\
$97.29$ & $-32.52$ & $8.6\times10^{-2}$ & \multicolumn{1}{c}{IC 5298} & Seyfert 2 galaxy \\
$98.88$ & $-36.55$ & $8.6\times10^{-2}$ & \multicolumn{1}{c}{NGC 7678} & active galaxy nucleus \\
$99.24$ & $-35.40$ & $8.6\times10^{-2}$ & \multicolumn{1}{c}{NGC 7673} & emission-line galaxy \\
$99.32$ & $-35.49$ & $8.6\times10^{-2}$ & \multicolumn{1}{c}{NGC 7677} & galaxy in pair of galaxies
\enddata
%% Include any \tablenotetext{key}{text}, \tablerefs{ref list},
%% or \tablecomments{text} between the \enddata and 
%% \end{deluxetable} commands
%% General table comment marker
%\tablecomments{\textcolor{black}{(a) The $\HI$ data is missing. (b) There are secondary $\HI$ peaks outside the velocity range from $-25\,\mathrm{km\,s^{-1}}$ to $+12\,\mathrm{km\,s^{-1}}$. (c) While this object is not listed in the \textit{IRAS} point source catalog, $\Td$ is high around the source.}}
\tablenotetext{(a)}{\textcolor{black}{The $\HI$ data is missing.}}
\tablenotetext{(b)}{\textcolor{black}{There are secondary $\HI$ peaks outside the velocity range from $-25\,\mathrm{km\,s^{-1}}$ to $+12\,\mathrm{km\,s^{-1}}$.}}
\tablenotetext{(c)}{\textcolor{black}{\citet{2011ApJ...733L..26A}}}
\tablenotetext{(d)}{\textcolor{black}{While this object is not listed in the \textit{IRAS} point source catalog, $\Td$ is locally high around this object.}}
\tablecomments{Columns 1 and 2 give the positions of each mask, and column 3 gives their areas. Column 4 indicates object names located at the center of each mask. Remarks on each masked area are listed in column 5. Details on these masked areas are described in Section 2.5.}
%% General table references marker
%\tablerefs{Table references}
\end{deluxetable}

\begin{deluxetable}{ccccc}
\tablewidth{0pt}
\tablecaption{Parameters of the data points in each $\Td$ range\label{table2}}
\tablenum{2}
\tablehead{\colhead{$\Td$} & \colhead{$N_{\mathrm{pixel}}$} & \colhead{Slope $k$} & \colhead{Intercept} & \colhead{C.C.} \\ 
\colhead{$(\mathrm{K})$} & \colhead{} & \colhead{$(\mathrm{K\,km\,s^{-1}})$} & \colhead{$(\mathrm{K\,km\,s^{-1}})$} & \colhead{} } 
\startdata
$17.0${--}$17.5$ & $1.48\times10^{3}$ & $6.83\times10^{6}$ & $1.5\phn\times10^{2}$ & 0.62 \\
$17.5${--}$18.0$ & $2.73\times10^{4}$ & $9.85\times10^{6}$ & $1.4\phn\times10^{2}$ & 0.72 \\
$18.0${--}$18.5$ & $8.64\times10^{4}$ & $1.41\times10^{7}$ & $1.2\phn\times10^{2}$ & 0.80 \\
$18.5${--}$19.0$ & $1.53\times10^{5}$ & $1.92\times10^{7}$ & $1.1\phn\times10^{2}$ & 0.81 \\
$19.0${--}$19.5$ & $1.61\times10^{5}$ & $2.50\times10^{7}$ & $0.96\times10^{2}$ & 0.86 \\
$19.5${--}$20.0$ & $1.42\times10^{5}$ & $3.24\times10^{7}$ & $0.88\times10^{2}$ & 0.88 \\
$20.0${--}$20.5$ & $9.52\times10^{4}$ & $4.03\times10^{7}$ & $0.77\times10^{2}$ & 0.86 \\
$20.5${--}$21.0$ & $4.33\times10^{4}$ & $4.08\times10^{7}$ & $0.81\times10^{2}$ & 0.79 \\
$21.0${--}$21.5$ & $1.13\times10^{4}$ & $3.54\times10^{7}$ & $1.0\phn\times10^{2}$ & 0.71 \\
$\geq21.5$ & $7.18\times10^{2}$ & $8.34\times10^{7}$ & \nodata & 0.62 \\
\enddata
\tablecomments{$N_{\mathrm{pixel}}$ is the number of pixels in each $\Td$ range. C.C. indicates the correlation coefficients for the points in each $\Td$ range. Columns 3 and 4 give values of the slope and intercept of the best fit linear relationship between $\WHImain$ and $\taudmain$ (Figure~\ref{tau353whi1}). The last row gives the slope assuming the intercept is zero in this case. The data are fitted by least-squares method.}
\end{deluxetable}

\end{document}